\documentclass[prb,a4paper,twocolumn,groupedaddress,english]{revtex4-1}
\usepackage{babel}
\usepackage{xcolor}
\usepackage{graphicx}
\usepackage{amsmath}
\usepackage{amssymb}

\definecolor{darkblue}{rgb}{0.1,0.2,0.6}
\definecolor{darkred}{rgb}{0.8,0.1,0.2}
\usepackage[colorlinks,citecolor=darkblue,linkcolor=darkred,urlcolor=darkblue]{hyperref} 
\usepackage[all]{hypcap} 

\makeatletter
\adddialect\l@English\l@english
\makeatother

\newcommand{\BigO}[1]{\ensuremath{\operatorname{O}\left(#1\right)}}
\newcommand{\bra}[1]{\langle\,#1\,|}
\newcommand{\ket}[1]{|\,#1\,\rangle}

\newcommand{\braket}[2]{\langle\,#1\, | \, #2\,\rangle}
\newcommand{\fullket}[1]{\left|#1\right>}
\newcommand{\fullbra}[1]{\left<#1\right|}

\newcommand{\E}{\mathrm{e}}
\newcommand{\D}{\mathrm{d}}

\newcommand{\co}{(Color online) }

\newcommand{\cf}{\textit{cf.} }
\newcommand{\ie}{\textit{i.e.} }
\newcommand{\eg}{\textit{e.g.} }

\newcommand{\up}{\uparrow}
\newcommand{\dw}{\downarrow}
\newcommand{\tr}{\mathrm{Tr}}

\definecolor{commentcolor}{rgb}{0.1,0.2,0.6}
\definecolor{commentcolor2}{rgb}{1,0,0}
\definecolor{todocolor}{rgb}{0.8,0.1,0.2}

\begin{document}
\title{Shannon-R\'enyi entropies and participation spectra across 3d $O(3)$ criticality}
\author{David J. Luitz}
\affiliation{Laboratoire de Physique Th\'eorique, IRSAMC, Universit\'e de Toulouse,
{CNRS, 31062 Toulouse, France}}
\email{luitz@irsamc.ups-tlse.fr}
\author{Fabien Alet}
\affiliation{Laboratoire de Physique Th\'eorique, IRSAMC, Universit\'e de Toulouse,
{CNRS, 31062 Toulouse, France}}
\author{Nicolas Laflorencie}
\affiliation{Laboratoire de Physique Th\'eorique, IRSAMC, Universit\'e de Toulouse,
{CNRS, 31062 Toulouse, France}}
\date{February 19, 2014}

\begin{abstract}
\makeatletter{}Universal features in the scalings of Shannon-R\'enyi entropies of many-body groundstates are
studied for interacting spin-$\frac{1}{2}$ systems across (2+1) dimensional $O(3)$ critical points, using quantum Monte Carlo simulations on dimerized and plaquettized Heisenberg models on
the square lattice. 
Considering both full systems and line shaped subsystems, $SU(2)$ symmetry breaking on the N\'eel ordered side of
the transition is characterized by the presence of a logarithmic
term in the scaling of Shannon-R\'enyi entropies, which is absent in the disordered gapped phase. 
Such a difference in the scalings allows to capture the quantum critical point using Shannon-R\'enyi entropies for
line shaped subsystems of length $L$ embedded in $L\times L$ tori, as the smaller
subsystem entropies are numerically accessible to much higher precision than for the full system.
Most interestingly, at the quantum phase transition an additive subleading constant $b_\infty^{*\rm line}=0.41(1)$ emerges in the
critical scaling of the line Shannon-R\'enyi entropy $S_\infty^\text{line}$. 
This number appears to be universal for 3d $O(3)$ criticality, as
confirmed for the finite-temperature transition in the 3d antiferromagnetic spin-$\frac{1}{2}$ Heisenberg model.
Additionally, the phases and phase transition can be detected in several features of the participation
spectrum, consisting of the diagonal elements of the reduced density matrix of the line subsystem.
In particular the N\'eel ordering
transition can be simply understood in the $\{S^z\}$ basis by a confinement mechanism of
ferromagnetic domain walls.
 
\end{abstract}
\maketitle
\makeatletter{}\section{Introduction}

How do coefficients of a wave-function change at continuous quantum phase transitions? In a given
basis, this question can be addressed by monitoring the behavior of (inverse)
participation ratios, which have a long history \eg in localization physics
\cite{wegner_inverse_1980,fyodorov_analytical_1992,evers_fluctuations_2000,evers_mirlin_rmp_08}. More recently, the study of related quantities such as Shannon-R\'enyi entropies (which
quantify the localization of the wave-function in a given basis) in {\it many-body} problems has
revealed an intriguing aspect: subleading terms in the finite-size scaling of these quantities
appear to carry universal information, characteristic of the physics contained in the ground-state
wave-function\cite{stephan_shannon_2009,stephan_renyi_2010,stephan_phase_2011,zaletel_logarithmic_2011,luitz_universal_2014}. For instance, they can characterize the
presence of broken continuous or discrete symmetry breaking in the ground-state, as well as
information on the universality class of continuous phase transitions. Most
previous studies\cite{stephan_shannon_2009,stephan_renyi_2010,zaletel_logarithmic_2011,stephan_phase_2011,atas_multifractality_2012,um_entanglement_2012,alcaraz_universal_2013}
on this topic focused on one-dimensional quantum systems, where both
analytical and numerical studies are easiest. In particular, dealing with the exponentially growing size of the Hilbert space of many-body problems, while maintaining a large enough total system size to study finite-size dependence, is a hurdle to surmount.  

Recently, we have introduced in Ref.~\onlinecite{luitz_universal_2014} convenient numerical methods to study the Shannon-R\'enyi entropies of many-body systems through a quantum Monte Carlo (QMC) sampling of the ground-state wave function.
This method allows studies of much larger systems than previously accessible in numerical
calculations, which is necessary for the analysis of universal behavior at continuous quantum phase
transitions where physical correlation lengths diverge. 

Here, we will study this problem for a non-trivial, yet well-understood quantum phase transition in
two-dimensional quantum magnetism: the transition between a N\'eel antiferromagnet and a quantum paramagnet in
two $S=1/2$ quantum spin Heisenberg models with varying antiferromagnetic couplings, namely
two-dimensional coupled dimers and plaquettes (see Fig.~\ref{fig:lattices}). The variation of the
ratio of two exchange couplings $g=J_2/J_1$ allows to couple isolated paramagnetic units (at $g=0$)
to form a two-dimensional antiferromagnet (at $g=1$) which spontaneously breaks $SU(2)$ symmetry at zero
temperature. A quantum critical point at $g_c$ in the 3d $O(3)$ universality
class\cite{troyer_97,matsumoto_ground-state_2001,wang_high-precision_2006,albuquerque_quantum_2008,wenzel_09,sandvik_computational_2010} separates the
quantum disordered and N\'eel ordered phases.
  
The first part of the paper (Sec.~\ref{sec:SR}) is devoted to the study of the behavior of subleading terms in the SR
entropies of the ground-state of Heisenberg magnets. Given a density matrix ${\hat \rho}$, the SR entropies are defined as:
\begin{equation}
    S_q = \frac{1}{1-q} \ln \sum_i \left(\rho_{ii}\right)^q, \quad \text{with} \quad \rho_{ii}=\bra{i} \hat{\rho} \ket{i},
    \label{eq:sr-entropies}
\end{equation}
where $\ket{i}$ are states of the computational basis in which SR entropies are calculated. Note
that the choice of the natural logarithm (base $\mathrm{e}$) fixes the units of SR entropies to ``nats''.

We will first consider in Sec.~\ref{sec:SRfull} the SR entropy of the full system composed of $N=L^2$ interacting $S=\frac{1}{2}$ spins on a square lattice, that is choosing $\hat\rho$ in Eq.~\eqref{eq:sr-entropies} to be the full  density matrix $\hat\rho=\fullket{\Psi} \fullbra{\Psi}$ of the ground state $\fullket{\Psi}$.
SR entropies are generally found to have a leading behavior which is
extensive\cite{atas_multifractality_2012,luitz_universal_2014} $S_q \sim a_q N$ where the
pre-factor $0 \leq a_q \leq \ln(2)$ (for spin $\frac{1}{2}$ systems) depends on details of the model ($J_2$ in that case). In the
$\{S^z\}$ basis considered throughout this work, we naturally expect $a_q$ to be `small' in the N\'eel
phase, and `large' in the quantum disordered phase. This is easily understood by considering the
limit $q=\infty$, where $S_\infty=-\ln(\max(\rho_{ii}))$. Here, $\max(\rho_{ii})=\max_{i}
|\braket{i}{\psi}|^2$ is the (modulus squared of the) maximal
coefficient of the groundstate wave-function expanded in the $\{S^z\}$ basis. For antiferromagnetic
systems, this is the coefficient of the N\'eel state $\ket{\hskip -0.2cm\uparrow \downarrow \uparrow \downarrow
\cdots}$, which is expected to be much larger in the antiferromagnetically ordered phase than in the disordered phase.
In the antiferromagnetic phase, the groundstate spontaneously breaks the continuous $SU(2)$ symmetry
and our previous work \cite{luitz_universal_2014} showed that this is reflected in a subleading
logarithmic correction: $S_q=a_qN +l_q \ln N +\cdots$. In the disordered phase, no symmetry is broken and
the subleading term is in general a universal constant (expected to be zero in the paramagnetic phase discussed in this paper).

For this first part of the paper (Sec.~\ref{sec:SR}), we will consider the case $q=\infty$ essentially for practical purposes. Indeed, $S_\infty$ is simpler to obtain numerically (within our QMC simulations) and more importantly, the leading term prefactor $a_\infty$ is smaller ($a_\infty < a_q$ for all finite $q$) which ensures that we can reach larger system sizes. Despite these facts, simulations of the SR entropy of full two dimensional systems are limited to relatively small sizes (up to $N=144$) close to the quantum phase transition, as the prefactor $a_\infty$ is still quite large in this region (see also the discussion in appendix \ref{sec:MC}).

To circumvent this, we next consider in Sec.~\ref{sec:SRline} the scaling of the SR entropy of a subsystem,
composed of a single line of size $L$ (the geometry of the subsystem is defined in Fig.
\ref{fig:lattices}) embedded in a periodic $L\times L$ torus.  
Subsystem SR entropies are defined in analogy to Eq.~(\ref{eq:sr-entropies}),
except that we now consider the {\it reduced} density matrix $\hat\rho_B$ of a subsystem $B$ which is obtained by performing a partial trace over the
rest of the system $A$: \begin{equation} \label{eq:rhob} \hat\rho_B = \tr_A \hat\rho \quad \Rightarrow \quad
\rho_{B,i_B i_B} = \sum_{j(i_B)} \rho_{j(i_B) j(i_B)}, \end{equation} with $\ket{j(i_B)} =
\ket{j_A}_A \otimes \ket{i_B}_B$, \ie where the basis state $\ket{j(i_B)}$ is a tensor product state
of subsystems basis states $\ket{j_A}_A$ and $\ket{i_B}_B$.

The rationale for choosing a line shaped subsystem is two-fold: first, we physically expect 
that the SR entropy of the
line also contains the information about antiferromagnetic ordering (since for instance the
correlation function $\langle S^z(0)S^z(r)\rangle$ along the line is defined in terms of diagonal
elements of the reduced density matrix). Second, the SR entropy of the line has a leading term
$S_q^{\rm line}=a_q^{\rm line} L$, and therefore takes much smaller values than for the full-system SR
entropy for the same value of $L$ ($L$ scaling versus $L^2$ scaling). This leads to a better
accuracy and allows to reach much larger linear sizes $L$ in our QMC simulations. 

Our results indicate that the SR entropy of the line also shows
a sub-leading $\ln(L)$ term in the N\'eel phase, and a constant term 
in the disordered phase which turns out to vanish (see below). Quite interestingly, the subleading term right at the quantum phase transition is
a constant $b_\infty^{*,{\rm line}}\neq 0$, which appears to be identical (within error bars) for the two models studied. This suggest that this constant is characteristic of the 3d $O(3)$
universality class to which both quantum phase transitions belong. Further simulations of the
finite-temperature ordering phase transition of the simple cubic $S=1/2$ Heisenberg model (also in the same $O(3)$ universality class) support this conjecture for antiferromagnetic interactions. We also expect a universal value $b_\infty^{*}$ for the full system, even though the limited
accuracy (due to the large value of $a_\infty$) of our simulations does not allow to prove this. 

For the verification of the universality of $b_\infty^{*,\text{line}}$, we have first performed extensive calculations of the spin stiffness in order to extract the best estimate for the value of the transition temperature, $T_c=0.94408(2)$, in agreement with Ref.~\onlinecite{Yasuda03}.

The SR entropies are global averages of all coefficients of the wave-function and their scaling with
the (sub-) system size thus capture correctly phases and phase transitions. It is interesting to
ask whether {\it each individual} coefficient (or reduced density matrix diagonal element in the
case of subsystems) also ``sees'' the quantum phase transition when $g$ is varied -- this
independently of their scaling with system size, as exemplified with the maximal diagonal entry of the (reduced)
density matrix, governing $S_\infty$ ($S_\infty^{\rm line}$). Motivated by this question, we study in the second
part of the paper (Sec.~\ref{sec:PS}) the behavior of each diagonal element of the reduced density
matrix $\rho^{{\rm line}}_{ii}$ for a line subsystem across the
transition. In analogy with the entanglement spectrum \cite{li_entanglement_2008,calabrese_entanglement_2008}, 
we define the ``participation spectrum'' as the set of pseudo-energies
$\xi_i^{\rm line} = -\ln{\rho^{\rm line}_{ii}}$. The participation spectrum develops into well-defined bands, which can be classified according to the magnetization and the number of ferromagnetic domain-walls separating segments having different N\'eel line configurations ($|\hskip -0.1cm\uparrow\downarrow\uparrow\downarrow\cdots\rangle$ or $|\hskip -0.1cm\downarrow\uparrow\downarrow\uparrow\cdots\rangle$. Identifying the lowest-lying states in this spectrum allows to understand the quantum phase transition in terms of an effective repulsion between such
domain-walls. Even though the participation spectra appear to differ at
first glance in the quantum disordered phases of the two studied models, we find that
this can be understood easily by classifying states according to the number of strong or weak domain
walls (this notion is dictated by the local physics of one of the two models considered). A striking
outcome of this analysis is that  \emph{all} individual levels (even
corresponding to assumedly irrelevant states such as the fully polarized state) harbor signs of the
quantum phase transition, as exemplified for instance by an inflection point (with respect to $g$)
for almost all $\xi_i^{\rm line}$. We analyze this in detail for the most probable state. Another interesting sign of the quantum phase transition is revealed by the study of the finite-size behavior of the width of the lowest-lying bands. 
We will finally conclude in Sec.~\ref{sec:conclusion} on the implications of our results while the
appendices contain specific details of the QMC procedure used (Appendix \ref{sec:MC}), as well as exact results in
the limit of $g\rightarrow 0$ for reference (Appendix \ref{sec:exact}).

Let us begin our paper (Sec.~\ref{sec:model}) by providing useful details on the models studied as well as on the finite-size scaling analysis.
 
\section{Models and methods of analysis}
\label{sec:model}

\begin{figure} 
    \includegraphics{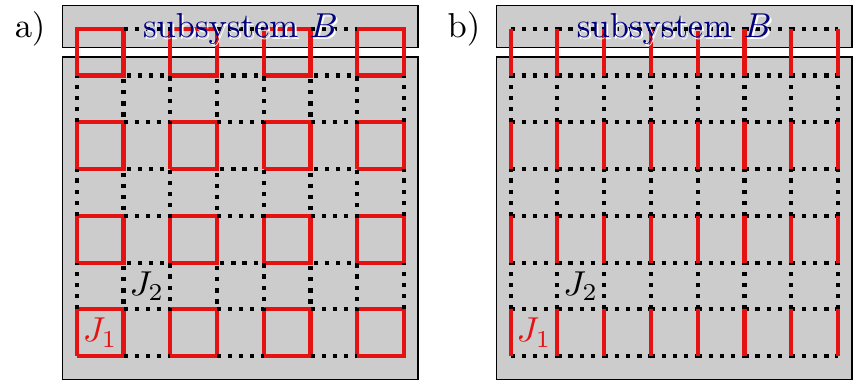}
    \caption{\co Plaquettized (left) and
        dimerized (right) lattices. The thick (red) lines correspond to strong bonds with coupling $J_1$ and we 
        refer to them as plaquettes/dimers.  The dotted bonds are the weak interplaquette/interdimer
    couplings $J_2\leq J_1$. Periodic boundary conditions are implicit. } 
    \label{fig:lattices}
\end{figure}

The two models (dimerized and plaquettized Heisenberg models) that we study are defined with the same Hamiltonian form:
\begin{equation} H= J_1 \sum_{\substack{\text{plaquettes/}\\ \text{dimers}}} \vec{S}_i \cdot
\vec{S}_j +J_2 \sum_{\text{links}} \vec{S}_i \cdot \vec{S}_j, \end{equation}
with $J_1,J_2\geq 0$ and where the two terms correspond to the summation over stronger bonds for columnar dimers (plaquettes)
and to the summation over the weaker links between these entities (see Fig.~\ref{fig:lattices}). We only consider
$g=J_2/J_1\leq1$ here, with $g=1$ yielding the homogeneous Heisenberg antiferromagnet on the square lattice.
The two models have slightly different critical points at
$g_c=0.52370(1)$\cite{sandvik_computational_2010} for the columnar dimerized system and
$g_c=0.54854(6)$\cite{wenzel_09} for the plaquettized system. For $g<g_c$ both models display a disordered ground state separated from excited states by a finite energy gap, whereas for $g>g_c$ antiferromagnetic N\'eel long-range order occurs, with a spontaneous breaking of the $SU(2)$ symmetry. 
We considered these two models as they are well-established to harbor the same physical content (in particular the quantum phase transitions at $g_c$ belong to the same 3d $O(3)$ universality class), yet with different microscopics: this will allow us to discuss universality of the scaling of SR entropies.

We study properties of the groundstates expanded in the $\{S^z\}$ basis and note that all
results will be identical in any basis obtained by a global $SU(2)$ transformation, by symmetry of
the Hamiltonian. We use the index $\square$ (respectively $|$) to denote quantities for the plaquettized (resp. dimerized) model. Considering the results of Ref.~\onlinecite{luitz_universal_2014}, we will perform fits of the SR entropy $S_\infty$ of the full system to the following forms:
\begin{equation}
    S_\infty(N)=a_\infty N + l_\infty \ln N + b_\infty 
    \label{eq:logscaling}
\end{equation}
and
\begin{equation}
    S_\infty(N)=\tilde{a}_\infty N + \tilde{b}_\infty.
    \label{eq:bscaling}
\end{equation}
Equivalent forms for the line SR entropy  $S_\infty^{\rm line}(L)$ are:
\begin{equation}
    S_\infty^{\rm line}(L)=a_\infty^{\rm line} L + l_\infty^{\rm line} \ln L + b_\infty^{\rm line} 
    \label{eq:logscalingline}
\end{equation}
and
\begin{equation}
    S_\infty^{\rm line}(L)=\tilde{a}_\infty^{\rm line} L + \tilde{b}_\infty^{\rm line}.
    \label{eq:bscalingline}
\end{equation}
Note that in general, one also expects~\cite{luitz_universal_2014} further size corrections  $\BigO{\frac{1}{N}}$ and  $\BigO{\frac{1}{L}}$.

The second functional forms Eqs.~\eqref{eq:bscaling} and \eqref{eq:bscalingline} are of course
included into the first forms Eqs.~\eqref{eq:logscaling} and \eqref{eq:logscalingline}, when the
fitting parameters $l_\infty$ or $l_\infty^{\rm line}$ are found to be zero. However, given the
finite values of $N$ and $L$ that we can reach and the error bars inherent to QMC, the fits to
Eqs.~\eqref{eq:bscaling} and \eqref{eq:bscalingline} are better controlled (and errors on estimated
parameters smaller) by forcing $l_\infty$ to be zero for systems where no log term is present.
Indeed, putting a log term when not needed can result in an acceptable fit where an artificial
$l_\infty > 0$ compensates wrongly underestimated $a_\infty$ or $b_\infty$. For systems where no log
term is present, we must have $\tilde{b}_\infty \rightarrow b_\infty$ and $\tilde{a}_\infty
\rightarrow a_\infty$ (respectively  $b^{\rm line} \rightarrow b_\infty^{\rm line}$ and
$\tilde{a}_\infty^{\rm line} \rightarrow a_\infty^{\rm line}$) for large enough sizes, but this
scaling regime might be reached earlier by using the second forms Eqs.~\eqref{eq:bscaling},
\eqref{eq:bscalingline}. Let us finally mention the simple argument that if one is looking for
universal constants, then only $l_\infty$ and $l_\infty^{\rm line}$ can be universal (but not $b_\infty,b_\infty^{\rm line}$) in the first forms Eqs.~\eqref{eq:logscaling} and\eqref{eq:logscalingline}: this is seen by a redefinition of sample size $N$ or $L$.  With the same reasoning, only $\tilde{b}_\infty,\tilde{b}_\infty^{\rm line}$ can be universal for the second forms Eqs.~\eqref{eq:bscaling} and \eqref{eq:bscalingline}.

For all fits, we used a rigorous bootstrap analysis in order to provide reliable error bars for fit parameters. Note, however, that these error bars do not contain systematic effects due to finite system sizes. These effects can nevertheless be estimated by comparison of fits over different system size $N$ or $L$ ranges (``fit windows'', see Ref.~\onlinecite{luitz_universal_2014} for details). We also monitored the fit quality $Q$ (see Ref.~\onlinecite{young_everything_2012}) to ascertain the precision of our fits.

\section{Shannon-R\'enyi entropies}
\label{sec:SR}
Throughout this section, we restrict our discussion and analysis to the computationally most accessible SR
entropy, when $q\rightarrow \infty$ for both the full system ($S_\infty$) in Sec.~\ref{sec:SRfull}
and the line subsystem ($S_\infty^{\rm line}$) in Sec.~\ref{sec:SRline}. 
\subsection{SR entropy $S_\infty$ of the full system}
\label{sec:SRfull}

\begin{figure} \includegraphics{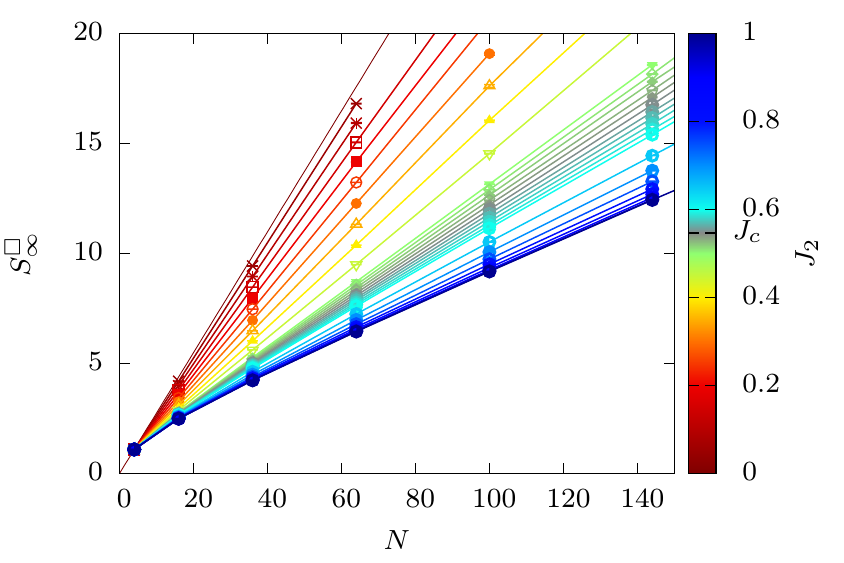} 
    \caption{\co $S_\infty^\square$ for different values of the plaquette coupling strength $J_2$.
        The result for the limit $J_2=0$ is given by $S_\infty^\square = \frac{\ln 3}{4} N$ (see appendix \ref{sec:exact}). An emerging logarithmic
        scaling term for $J_2>J_c$ can be guessed.
        Lines are guides to the eye.
    } 
    \label{fig:plaq_full_sinf}
\end{figure}

Figure \ref{fig:plaq_full_sinf} shows our QMC result for $S^\square_\infty$ of the
plaquettized model in the range of accessible entropies (our simulations are limited roughly to  $S_\infty \lesssim 20$ as
discussed in appendix \ref{sec:MC}), for different values of the parameter $J_2$ in the range $[0,1]$. 

In the limit $J_2=0$ of isolated plaquettes, $S_\infty^\square(N)$ can be exactly
(\cf appendix \ref{sec:exact}) shown to be $S_\infty^\square(N)=\frac{\ln 3}{4} N$, {\it i.e.} a pure linear scaling with no logarithmic or constant terms. In the uniform Heisenberg limit $J_2=J_1$ on the other hand, previous results~\cite{luitz_universal_2014} have shown the existence of a logarithmic scaling correction with the form Eq.~\eqref{eq:logscaling} with $l_\infty \neq 0$.

By inspection of the bare SR entropy scaling in Fig. \ref{fig:plaq_full_sinf}, a nonzero logarithmic
scaling term $l_\infty>0$ can be presumed for the whole ordered phase $J_2>J_c$ (with curves clearly
bending downwards for smaller system sizes), while for the disordered phase, the scaling appears
linear. In order to quantify this, we have performed fits of the Monte Carlo data corresponding to
the form Eq. (\ref{eq:logscaling}). We emphasize that the quality of the fits (in particular the
extraction of the logarithmic term) is reduced when only few system sizes are available, which is
specially the case in the disordered regime of the phase diagram (due to faster growing
$S_\infty$ with system size). As the situation is worse for the dimerized model (we have for
instance $a_\infty^|(J_2=0)=\frac{\ln 2}{2} > a_\infty^\square(J_2=0) = \frac{\ln 3}{4}$, \cf
appendix \ref{sec:exact}), we concentrated our analysis for this section on the plaquettized model.

\begin{figure} \includegraphics{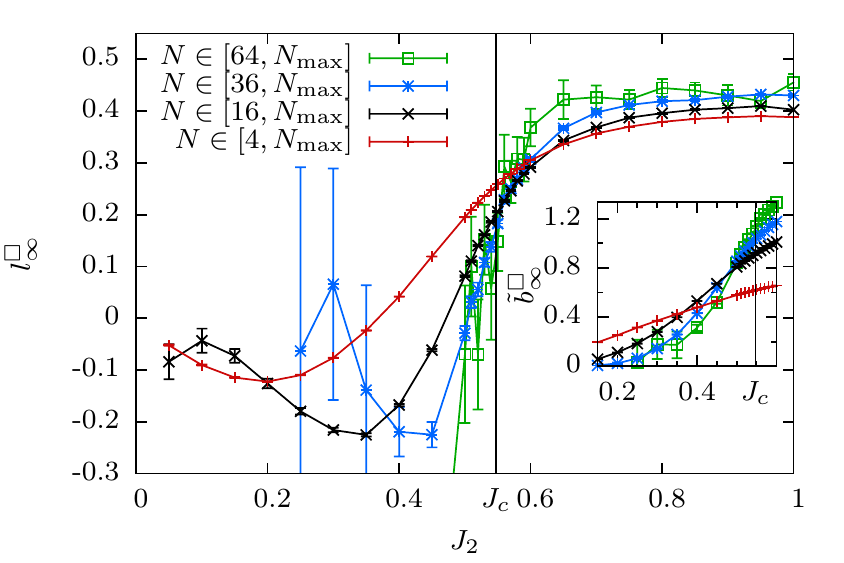} \caption{\co Prefactor
        $l_\infty^{\square}$ of the
        logarithmic scaling term of $S_\infty^\square$ extracted from fits over different size windows for the
    plaquettized square lattice. $N_{\rm max}$ is the maximal accessible size for which the entropy
    is smaller than  $\approx 20$ (see Fig.~\ref{fig:plaq_full_sinf}). We used $N_{\rm max}=144$ for
    $J_2>0.3$ and were able to push calculations up to $N_{\rm max}=196$ around the critical point
    and even to $N_{\rm max}=256$ for $J_2=1$.
    For $J_2<J_c$, the fits are difficult because of greater errorbars
for large entropies and large finite size effects. The behavior is nevertheless consistent with a
vanishing $l_\infty^\square$ in the disordered phase. For $J_2>J_c$ a plateau emerges and
$l_\infty^\square$ is found to assume approximately the same nonzero value in the whole ordered phase.
The inset shows the subleading constant term $\tilde{b}_\infty^\square$ obtained from fits excluding a
logarithmic scaling term, in the relevant low-$J_2$ phase. 
}
\label{fig:plaq_linf}
\end{figure}

Fig. \ref{fig:plaq_linf} displays the result of our fits for the prefactor $l_\infty^{\square}$ of
the logarithmic scaling correction of $S_\infty^\square$. The trend with increasing system sizes
included in the fit is evident in the ordered phase, as $l_\infty^\square$ is found to be almost constant with $J_2$
there. In the disordered phase, large finite size effects are observed which are very similar to the
oscillations found for the constant term close to the quantum phase transitions of transverse field Ising models
\cite{stephan_shannon_2009,luitz_universal_2014}. We expect $l_\infty^\square$ to vanish in the
complete quantum disordered phase (as it is shown analytically for $J_2=0$ in appendix
\ref{sec:exact}) and our data are consistent with this expectation, although the numerical
precision is not sufficient for a definite answer. The lack of availability of larger $N$ also
prevents us to conclude if there is a universal number $l_\infty$ (and what is its numerical value)
in the N\'eel phase, even though the plateau shape of the curves tend to indicate that this is
possible. The actual universal value $l_\infty$ (if any) may be quite larger than the maximum value
here (found to be $l_\infty \simeq 0.45$ for the fit window with the largest $N$), as can be seen by the shift of the curves when smaller sizes are removed from the fit. 

The inset of Fig. \ref{fig:plaq_linf} shows our fit results for the same fit windows as in the main
panel for $\tilde{b}_\infty^\square$ as obtained from fits to Eq. \eqref{eq:bscaling} close to
$J_c$. In this region, large finite size effects are hampering a reliable extraction of the constant
but a lower bound for the value $b_\infty^{*,\square}\gtrsim 1.1$ at the critical point  can be
perceived. Results from fit windows excluding smaller system sizes seem to indicate that
$b_\infty^\square$ vanishes in the disordered phase.

It would be of clear interest to increase the maximum size in the simulation to have a larger
fitting range, but this is not possible with the extensive growth of the entropy $S_\infty$. To
circumvent this problem, we consider in the next section the scaling behavior of a the SR entropy of a subsystem, which grows much more slowly.

\subsection{SR entropy $S_\infty^{\rm line}$ of a line subsystem}
\label{sec:SRline}

\begin{figure}[h]
    \centering
    \includegraphics{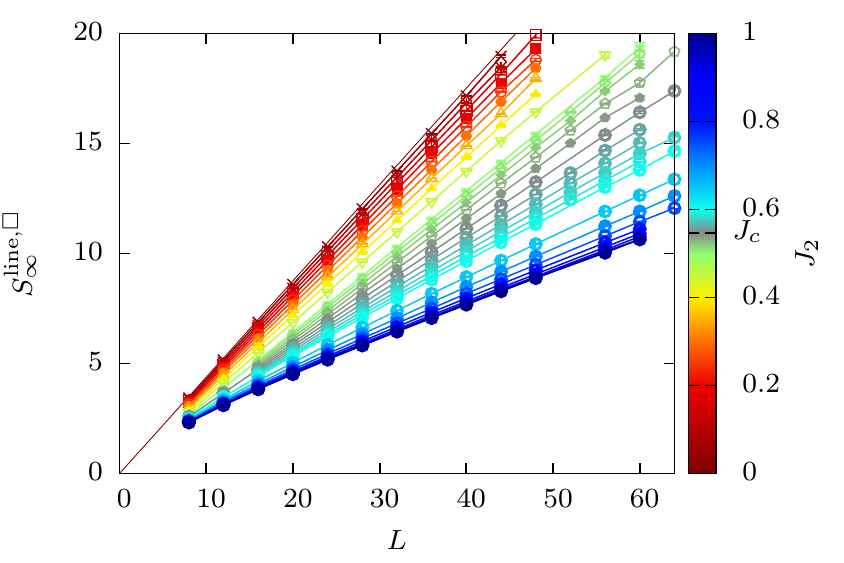}
    \includegraphics{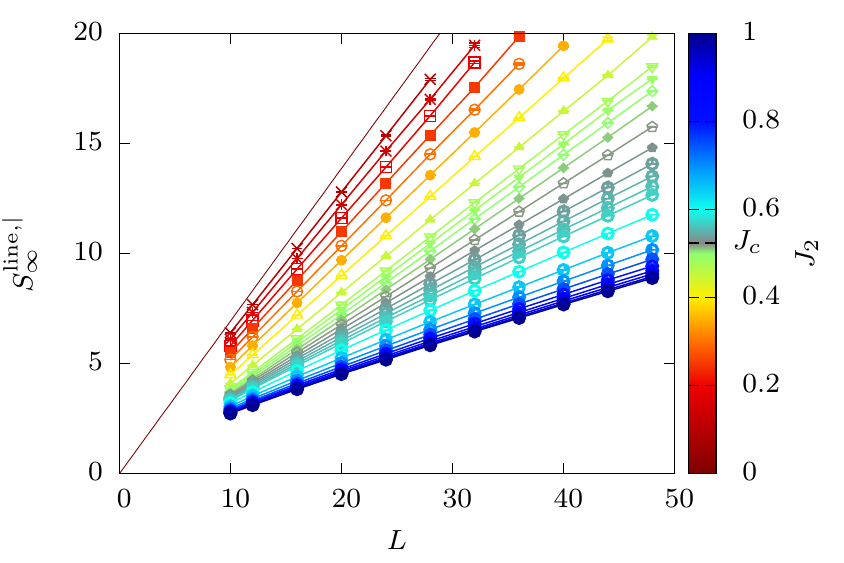}
    \caption{\co SR entropies for the line-shaped subsystem across the
plaquettization  ($S_\infty^{\text{line},\square}$, top panel) and dimerization ($S_\infty^{\text{line},|}$, bottom panel)  transitions.}
    \label{fig:all_line_sinf}
\end{figure}

We present in this section our QMC results for the line subsystem SR entropy  $S^{\rm line}_\infty$.
Its scaling with the length of the line $L$ will be shown to also capture the nature of the ordered
and paramagnetic phases. $S^{\rm line}_\infty$ is equal to (minus) the natural logarithm of the maximum diagonal entry of the line reduced density matrix, which turns out to correspond to the two local N\'eel states $| \hskip -0.1cm\uparrow \downarrow \uparrow \downarrow \cdots \rangle $ and $|\hskip -0.1cm \downarrow \uparrow \downarrow \uparrow\cdots \rangle $ on the line. This is slightly less obvious than the fact that the full N\'eel states are the most probable states on the full lattice, but we checked explicitly that this is the case in all our simulations. By definition of the reduced density matrix, $S_\infty^{\rm line} = -\ln (\max_i \rho_{ii,B})$ contains now information about all basis states of the full system which fulfill the geometrical condition of forming one of the two N\'eel states on the subsystem.

\begin{figure} 
\begin{center} 
\includegraphics{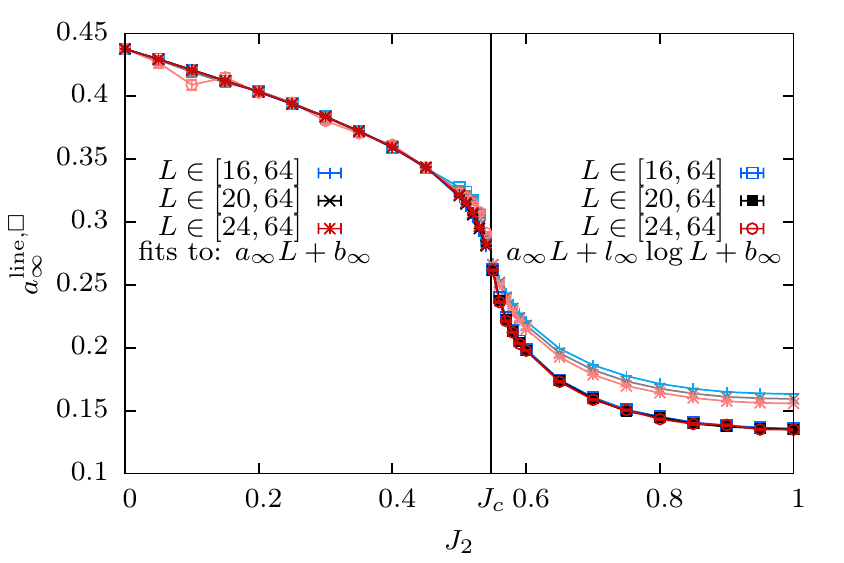} 
\includegraphics{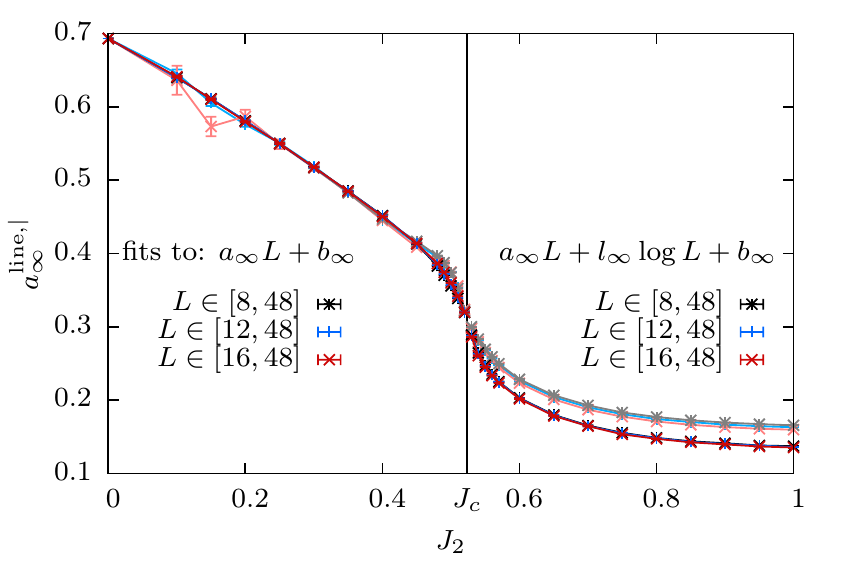} 
\end{center}
\caption{\co  Linear scaling prefactors $a_\infty^{\rm line}$ of the subsystem entropy
    $S_\infty^{\text{line}}$ across the plaquettization (top) and dimerization (bottom) transitions. 
    We show both fits to the forms $\tilde{a}_\infty^{\rm line} L + \tilde{b}_\infty^{\rm line}$
    (Eq.~(\ref{eq:bscalingline}), valid for $J_2<J_c$, bold in valid regime, pale for $J_2>J_c$)
    and $a_\infty^{\rm line} L + l_\infty^{\rm line} \ln L + b_\infty^{\rm line}$ (Eq.
    (\ref{eq:logscalingline}), valid for $J_2>J_c$, bold in valid regime, pale for $J_2<J_c$). As
    $b_\infty^{\rm line}=0$ and $l_\infty^{\rm line}=0$ in the quantum disordered phase when $J_2<J_c$, the fits forcing
    $l_\infty^{\rm line}=0$ are slightly better. In the ordered phase, the fit to Eq.~\eqref{eq:bscalingline} does
    not work because of the existence of the logarithmic scaling term and fit quality factors of
    $Q\approx 0$ (see \eg Ref. \onlinecite{young_everything_2012}) were obtained here. }
\label{fig:all_line_ainf}
\end{figure}

We display our results for the line subsystem SR entropies as a function of the length $L$ of the
subsystem for both dimerized and plaquettized models in Fig.~\ref{fig:all_line_sinf}. Much larger system sizes $N=L^2$ are accessible now (when compared to Fig.~\ref{fig:plaq_full_sinf} for the full system): this greatly reduces the effect of further finite size corrections beyond Eqs.~\eqref{eq:logscalingline} and \eqref{eq:bscalingline} and makes a reasonable analysis of the scaling of subsystem entropies viable. We now discuss systematically the scaling behavior of the SR entropy $S_\infty^\text{line}$
across the plaquettization-dimerization transitions, by fitting to the functional forms Eq.~\eqref{eq:logscalingline} and \eqref{eq:bscalingline}, and displaying the estimates of fits parameters. 

\begin{figure} 
\begin{center} 
\includegraphics{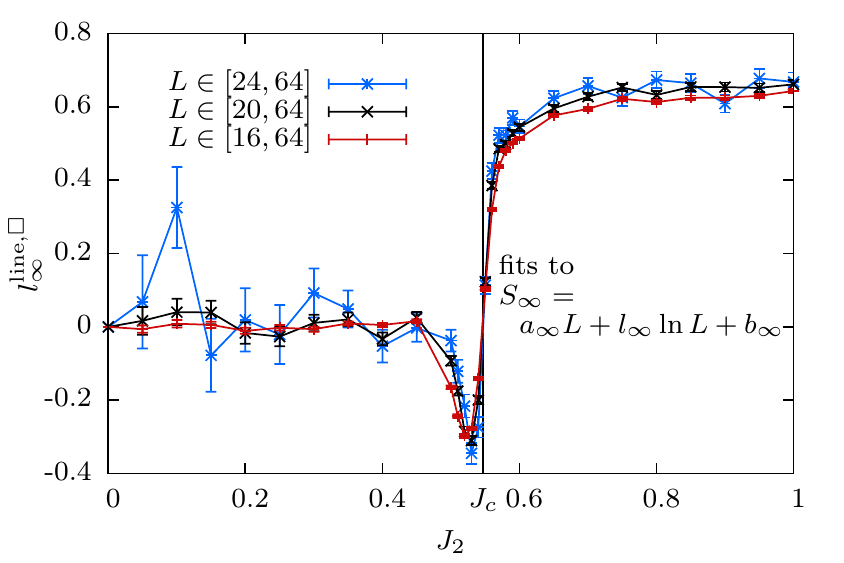} 
\includegraphics{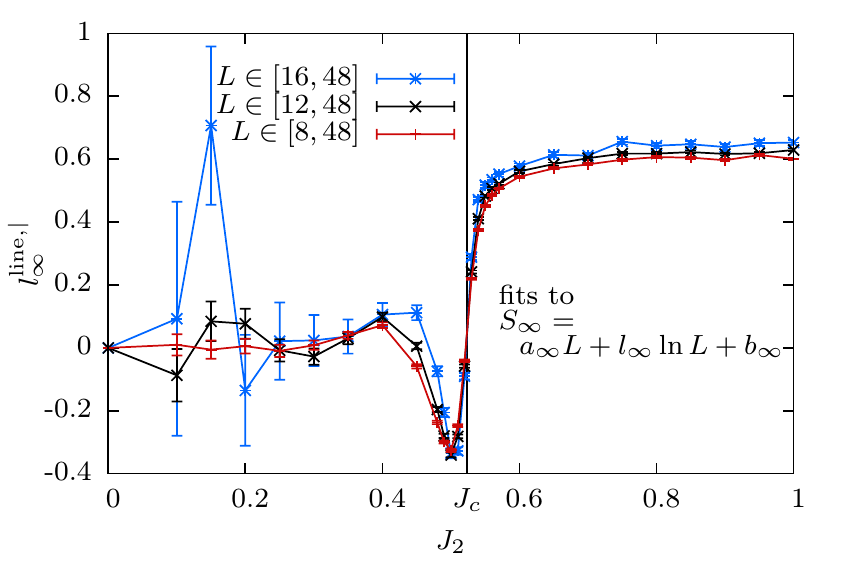}
\end{center}
\caption{\co Logarithmic scaling term $l_\infty^{\rm line}$ of the line SR entropy $S_\infty^{\rm line}$ across the
plaquettization (top) and dimerization (bottom) transitions, as obtained from fits to Eq.~\eqref{eq:logscalingline}.  We show fits over different system size windows. The logarithmic term
vanishes in the quantum disordered phase, while in the ordered phase it assumes a nonzero, almost constant value, which is similar for both models for a given fitting size window. } 
\label{fig:all_line_linf}
\end{figure}

\begin{figure} 
\begin{center}
 \includegraphics{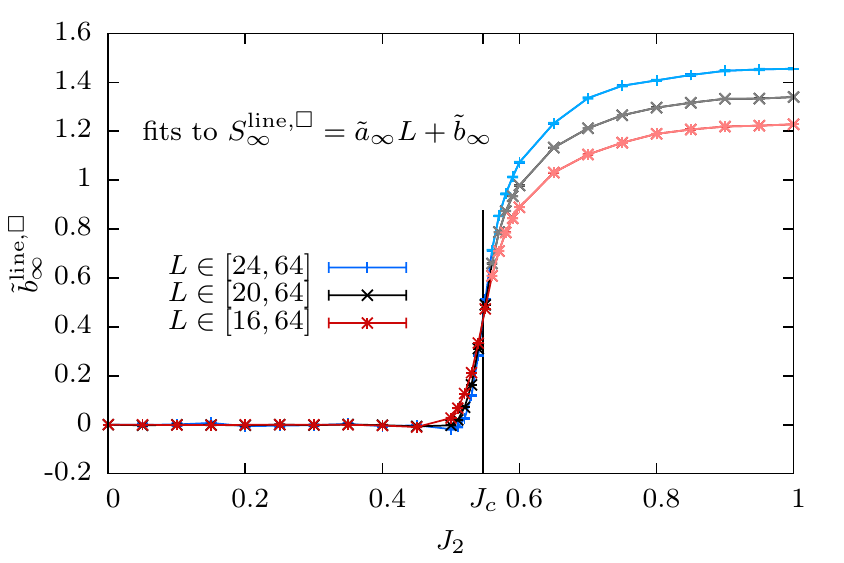} 
 \includegraphics{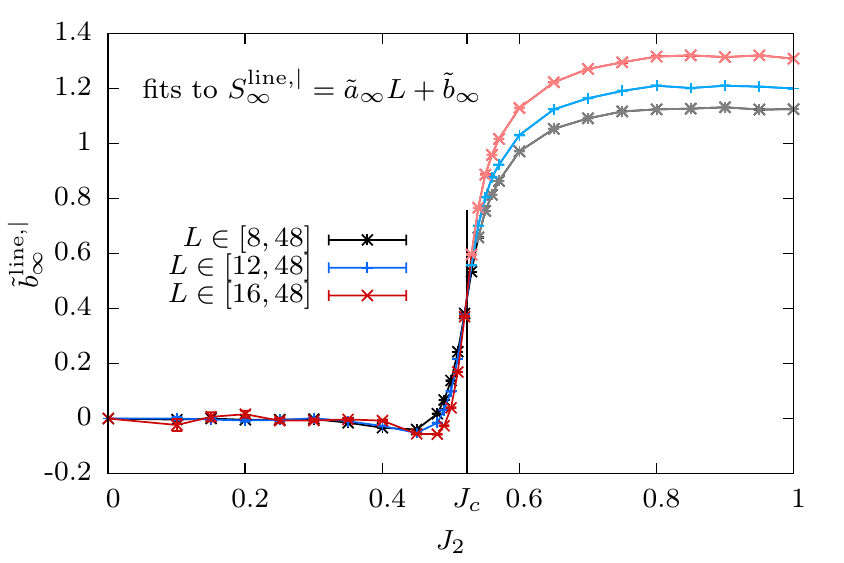} 
  \end{center}
    \caption{\co Constant scaling term $\tilde{b}_\infty^{\text{line}}$ of the subsystem entropy $S_\infty^{\rm line}$ across the
plaquettization (top) and dimerization (bottom) transitions, as obtained by a fit to Eq.~\eqref{eq:bscalingline}. This form is
clearly not valid in the ordered phase ($J_2>J_c$), where a logarithmic scaling term of $l_\infty^{\rm line}>0$
is found. Fit qualities drop to zero for $J_2>J_c$ and $\tilde{b}^{\text{line}}_\infty $ is
therefore shown in pale colors. In the disordered phase $J_2<J_c$,
$\tilde{b}_\infty^{\text{line}} ={b}_\infty^{\text{line}}$ is found to be 0 (bold).
The lines cross at the critical point at $b_\infty^{*,\text{line},\square} = 0.412(6)$ (plaquettized model) and $b_\infty^{*,\text{line},|} = 0.41(1)$ (dimerized model).
 } 
\label{fig:all_line_binf}
 \end{figure}

\subsubsection{Leading term}
We begin with the linear prefactors $a_\infty^{\text{line}}$, as displayed in Fig. \ref{fig:all_line_ainf} as a
function of $J_2$ for fits over different system size windows. The results for the two functional
forms are shown in the same figure, but with a different color coding depending on the regimes:
$a_\infty^{\rm line}$ obtained from linear fits [Eq.~\eqref{eq:bscalingline}] is represented with
bold lines for $J_2<J_c$ (in the disordered regime where we find that they represent the correct form) and pale lines for $J_2>J_c$ (when they are not expected to be valid) and vice-versa for fits including the logarithmic correction [Eq.~\eqref{eq:logscalingline}]. For $J_2<J_c$, both results agree very well within error bars, while the linear fit result is slightly more stable and converges faster with system size. This is already a hint that the logarithmic correction $l_\infty^{\text{line},\square}$ presumably vanishes in the disordered phase, which will be verified in the next paragraph. Both dimerized and plaquettized models display the same behavior, with $a_\infty^{\text{line},|}$ taking larger values due to suppressed N\'eel order. 

One can notice a qualitative change in the extensive contribution to the Shannon entropy across the quantum phase transition where $a_{\infty}$ changes abruptly. More precisely, its derivative with respect to $J_2$ displays a singularity at the critical point. We discuss in more detail such features in Sec.~\ref{sec:detecting}.

\subsubsection{Subleading logarithmic term in the ordered phase}
The first subleading scaling term is the logarithmic correction $l_\infty^{\text{line}}$ as defined
in Eq. \eqref{eq:logscalingline}. Fig.~\ref{fig:all_line_linf} represents results of fits obtained from three sets of system size ranges. We find that fits excluding the smallest system sizes generally correspond to higher fit qualities (quality factor $Q$ closer to $1$) while on the other hand, error bars on $l_\infty^{\text{line}}$ become larger as the
number of data points included in the fit decreases. 

Nevertheless, results are stable with respect to different fit windows: we observe a clear
change in the estimated $l_\infty^{\text{line}}$ exactly at the transition point for both dimerized and
plaquettized models at the respective $J_c$. Deep in the quantum disordered phase, the logarithmic term
$l_\infty^\text{line}$ converges very well towards zero. Close to the critical point for $J_2<J_c$,
nontrivial finite size effects show up in pronounced oscillations preceding the jump to nonzero
$l_\infty^{\text{line}}$ in the ordered phase. Similar to what is observed in the constant term of
the SR entropies of the one-dimensional~\cite{stephan_shannon_2009} and two-dimensional~\cite{luitz_universal_2014} quantum Ising model close to its transition point, the oscillations become narrower and move closer to the critical point with growing system sizes used for the fit. We conclude that $l_\infty^{\text{line}}=0$ in the full disordered phase.

In the ordered phase, the behavior is very different and a logarithmic scaling correction emerges
with $l_\infty^\text{line}>0$. Our results for the fitting window with the larger sizes is $l_\infty^\text{line} \gtrsim 0.7$ and
$l_\infty^\text{line}$ appear identical for both models within the N\'eel phase. However, even though we performed
calculations in large systems of up to $N=4096$ spins, the asymptotic value of $l_\infty^\text{line}$ cannot be
extrapolated from our data.

Right at the critical point, curves for the estimated $l_\infty^\text{line}$ for different fit windows cross at a value which is $0$ within error bars. 
 
\subsubsection{Vanishing constant term in the paramagnetic phase}
In the quantum disordered phase and presumably also at the critical point, the logarithmic term vanishes and therefore the first subleading scaling term is $b_\infty^\text{line}$. To best estimate its value, we force $l_\infty^\text{line}=0$ by using the functional form Eq.~\eqref{eq:bscalingline} in our fit. Fig.~\ref{fig:all_line_binf} shows the result of this
analysis, the pale lines correspond to the regime $J_2>J_c$ where the fit function does not represent the data
correctly (this is reflected also by strong finite size effects). We find exactly the same behavior
for both models in the disordered phase with $b_\infty^\text{line} = 0$ for all $J_2<J_c$. 

\subsubsection{Universal constant term at the quantum phase transition}
We furthermore find (see Fig.~\ref{fig:all_line_binf}) that curves of $\tilde{b}_\infty^\text{line}$ for
different fit windows cross at the critical point, taking a non-trivial value
$b_\infty^{*,\text{line}}$. The absence of finite size effects at the crossing point provides
evidence that the logarithmic correction actually vanishes at the critical point.
For the plaquettized model, we find $b_\infty^{*,\text{line},\square} = 0.412(6)$; in the dimerized case we obtain a similar value
$b_\infty^{*,\text{line},|} = 0.41(1)$. This strongly suggests that $b_\infty^{*,\text{line}}$ is {\it universal} at the quantum critical point, and should be identical for all models with a phase transition in the 3d $O(3)$ universality class.

To test this, we perform large-scale simulations of the {\it finite-temperature} transition in the isotropic 3d $S=1/2$ Heisenberg model on a cubic lattice with antiferromagnetic interactions. This transition belongs to the 3d $O(3)$ universality class. We then computed the line SR entropy $S_\infty^{\text{line}}$ using the same QMC technique~\cite{luitz_universal_2014}, but this time at finite temperature, close to the critical point.

\begin{figure}
    \centering
    \includegraphics{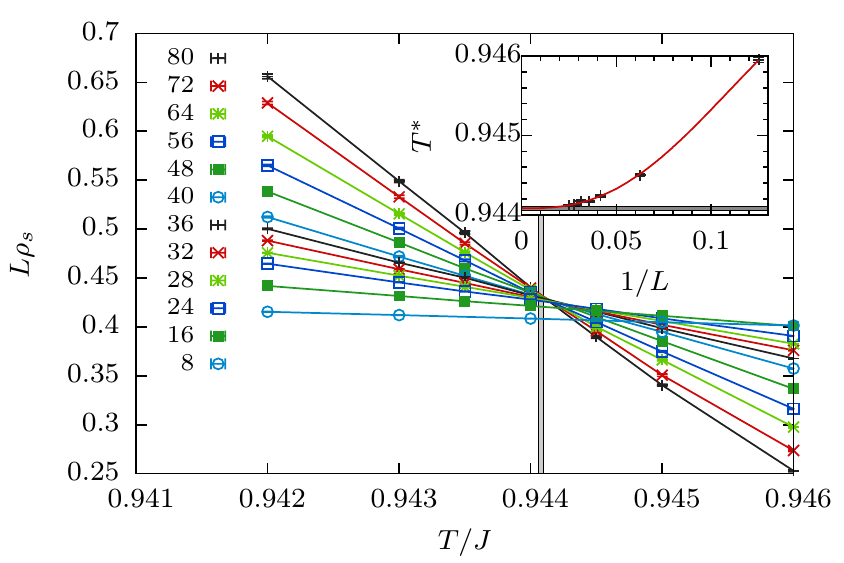}
    \caption{Spin stiffness multiplied by system size $\rho_s L$ as a function of temperature $T$ for different linear system
    sizes $L$ in the 3d $S=1/2$ antiferromagnetic Heisenberg model on the simple cubic lattice.
    The inset shows our best estimate for the crossing point $T^*$ of the spin stiffness for sizes
    $L$ and $2L$ as obtained from a cubic fit to our data including a bootstrap analysis for the
    error bars (black points). The red line corresponds to $T^*$ calculated from our best fit of our
    data to a universal function $f(L,T) = (1+c/L^\omega) k( L^{1/\nu}(T-T_c) + d/L^{\phi} )$
    (\cf Ref. \onlinecite{wang_high-precision_2006}),
    including empirical scaling corrections. We approximated $k$ by a third order polynomial.
    For the critical temperature, we obtain $T_c=0.94408 \pm 0.00002$ using $10^4$ bootstrap
    samples (1$\sigma$ errorbar indicated by the narrow rectangle).
}
    \label{fig:3dTc}
\end{figure}

\begin{figure}
    \centering
    \includegraphics{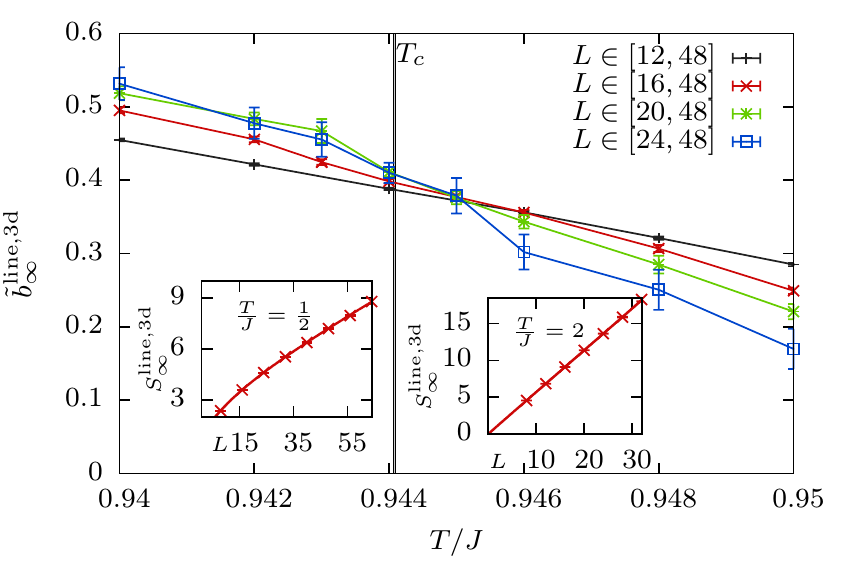}
    \caption{Fit result for the subleading constant $b_\infty^{\text{line,3d}}$ in the scaling 
    of the SR entropy $S_\infty^{\text{line,3d}}$ with system size close to the critical temperature
    in the 3d Heisenberg model. We only performed fits to the form in Eq. (\ref{eq:bscalingline}),
    which is strictly only valid in the absence of logarithmic scaling terms (\ie in the disordered
    phase at $T\geq T_c$). At the critical point, $b_\infty^{\text{line,3d}}$ converges well with
    system size and the estimate for the fit window with the largest sizes is given by
    $b_\infty^{*,\text{line,3d}}=0.41(1)$. The insets show the scaling of $S_\infty^{\text{line,3d}}$ as a function
    of $L$ in the ordered phase $T<T_c$ with a clear sign of the logarithmic scaling correction
    ($l_\infty^{\text{line,3d}}=0.8(3)$) and
    in the paramagnetic phase $T>T_c$, where the scaling is purely linear with a vanishing constant
    $b_\infty^{\text{line,3d}}=0.003(9)$.
}
    \label{fig:3db}
\end{figure}

As a preliminary, we want first to extract the best estimate for the critical temperature $T_c$.
We have performed additional simulations, up to $N=512000$ sites, studying the crossings of the spin stiffness (times linear system size), a standard method to locate critical points~\cite{sandvik_computational_2010}.
These results are reported in Fig.~\ref{fig:3dTc} where we show very precise QMC data for cubic
systems, thus allowing to estimate the critical point with a high accuracy to
$T_c/J=0.94408(2)$, which agrees with previous estimate $T_c/J=0.944175 (75)$ from Ref.~\onlinecite{Yasuda03}. While measuring the spin stiffness within the SSE computation is very
standard~\cite{sandvik_computational_2010} and relatively fast, accessing $S_\infty$ for a single line in the cubic
antiferromagnet requires much longer simulation time. We have  been able to reach system
sizes up to $N=48^3$ for $S_{\infty}^{\rm line, 3d}$ for which the subleading constant $b_\infty^{\text{line,3d}}$ is
shown in Fig.~\ref{fig:3db} in the vicinity of $T_c$. Despite the sizable error bars, we can nevertheless observe a clear crossing for various fit windows,
drifting towards the actual critical point at $T_c/J=0.94408(2)$ where the subleading constant takes
a numerical value $b_\infty^{\text{line,3d,*}}=0.41(1)$. This value is in perfect agreement with
estimates for the two-dimensional quantum critical points, thus reinforcing the evidence for the universality of
$b_\infty^{\text{*, line}}=0.41(1)$ for 3d O(3) critical points.

Away from criticality, we have also checked the scalings of $S_{\infty}^{\rm line, 3d}$ in the low temperature ordered phase at $T=J/2<T_c$ (left inset of Fig.~\ref{fig:3db}) and in the high temperature disordered regime at $T=2J>T_c$ (right inset of Fig.~\ref{fig:3db}). 
As expected, below $T_c$, a subleading logarithmic term emerges with $l_\infty^{\text{line,3d}}=0.8(3)$, and a purely linear scaling is found above $T_c$, with a vanishing constant $b_\infty^{\text{line,3d}}=0.003(9)$.  

\section{Participation spectra}
\label{sec:PS}

Up to now, we have focused on the finite-size behavior of a single quantity, namely $S_\infty$ ($S_\infty^{\rm line}$), which is related to a single diagonal element - the largest - of the (reduced) density matrix.  Let us now inspect the behavior of all the diagonal elements of the reduced density matrix $\hat \rho_B$ for a subsystem $B$ being, as above, a line of $L$ spins embedded in a $L\times L$ torus. For practical reasons, we again restrict ourselves to the set of bases
that are connected to the $\{S^z\}$ basis by global $SU(2)$ transformations, leaving the Hamiltonian
invariant. 

\subsection{Definitions}

Inspired by recent insights obtained on the \emph{entanglement spectrum}~\cite{li_entanglement_2008,calabrese_entanglement_2008,stephan_renyi_2012}, we introduce the
\emph{participation spectrum} obtained from the diagonal of the reduced density matrix $\hat\rho^{\rm line}$ in
the computational basis $\{ \ket{i} \}$ \begin{equation} \xi_i^{\rm line} = -\ln{\rho}_{ii}^{\rm line} =-\ln \left( \bra{i} \hat{\rho}^{\rm line} \ket{i}
\right),  \end{equation} 
using the line shaped subsystem defined in Fig. \ref{fig:lattices}. From now on, we drop the index `line' on the set of pseudo-energies $\xi_i$.

In order to clarify the tremendous amount of information contained in the participation spectrum, we anticipate (as detailed below) that the line participation spectrum will develop well-defined bands that can be classified through specific characteristics of their containing basis states:  (absolute value of) magnetization $|S^z|$ and number of (ferromagnetic) domain-walls $n_{\rm dw}$, the later turning out to be the crucial element to classify the spectrum.

The $S^z$ operator being diagonal in the computational basis $\{ \ket{i} \}$, the magnetization of a
basis state $\ket{i}$ is simply defined as $S^z(\ket{i})=\bra{i} S^z \ket{i} $. We define the total
number of ferromagnetic domain walls in the line as $n_{\rm dw}(\ket{i})=L/2+2\sum_{x=1}^{L}
\bra{i} S^z_x S^z_{x+1} \ket{i}$. We assume periodic boundary conditions ($S^z_{L+1}=S^z_1$) along
the chain with $L$ even (ensuring $n_{\rm dw}$ to be an even number). In other words, $n_{\rm dw}$
is simply the number of bonds along the chain hosting nearest-neighbors spins with the same
orientation in basis state $\ket{i}$.   We use the term {\it domain walls} since the most likely
states are the two N\'eel states on the chain, as mentioned earlier, which have $|S^z|=0$ and $n_{\rm
dw}=0$  (ferromagnetic orientation of spins on a given bond correspond to disrupting local N\'eel
ordering). Magnetization $|S^z|$ and number of domain walls $n_{\rm dw}$ run from $0$ to
respectively $L/2$ and $L$ (for the polarized ferromagnetic state) for possible basis states on the
chain. We expect states with low $n_{\rm dw}$ and low $|S^z|$ to be more likely, and therefore to have a lower pseudo-energy $\xi_i$. 

\begin{figure}
    \centering

\includegraphics[width=\columnwidth]{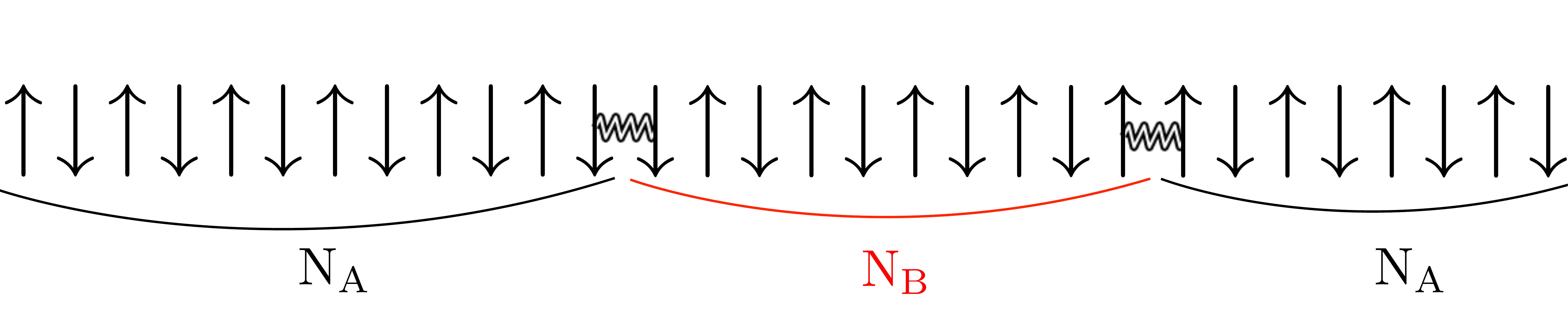}
    \caption{\co Schematic picture for a basis state having two domain walls $\downarrow\downarrow$ and $\uparrow\uparrow$ separating the two N\'eel configurations $\rm N_A$ and $\rm N_B$.
}
    \label{fig:dwspict}
\end{figure}

A typical basis state is illustrated in Fig.~\ref{fig:dwspict} for a line of 30 spins with $S^z=0$
and $n_{\rm dw}=2$. From such a picture one sees that increasing the separation between two domain
walls tends to reduce the total staggered magnetization $m_{\rm stag}=\sum_x (-1)^x \langle S_x^z\rangle$. It is therefore expected that the effective interaction between domain walls will be strongly (weakly) attractive for states having long-range (short-range) antiferromagnetic correlations. This will be discussed on more quantitative grounds below in Sec.~\ref{sec:detecting}.

We finally note that this description is not sufficient for characterizing states of the chain subsystem in the plaquettized lattice, as readily seen in Fig.~\ref{fig:lattices}. While all bonds along the chain are equivalent for the dimerized lattice, this is not the case for the plaquettized lattice with ``strong'' bonds carrying the coupling constant $J_1$  and ``weak'' bonds carrying $J_2\leq J_1$. We therefore find it useful to define the number of strong  $n_{\rm strong}(\ket{i})=L/4+2\sum_{s=1}^{L/2}  \bra{i} S^z_{2s} S^z_{2s-1} \ket{i}$  domain walls ($0 \leq n_{\rm strong} \leq n_{\rm dw}$). We assume that the chain subsystem starts from a strong bond and that $L$ is a multiple of $4$, as is the case in Fig.~\ref{fig:lattices}. We  expect that ferromagnetic domain-walls on strong bonds will be less likely than on weak bonds and anticipate this notion of strong and weak domain walls to be particularly relevant in the quantum disordered phase.

\subsection{Participation density of states}

To get a first idea on how the weight of each basis state gets redistributed while passing
through the quantum phase transition, it is instructive to consider the density of states \begin{equation}
\text{DOS}(\omega)= \frac{1}{2^L} \sum_i \delta(\omega - \xi_i) \end{equation}
corresponding to the participation spectrum of the line subsystem.

\begin{figure} \includegraphics{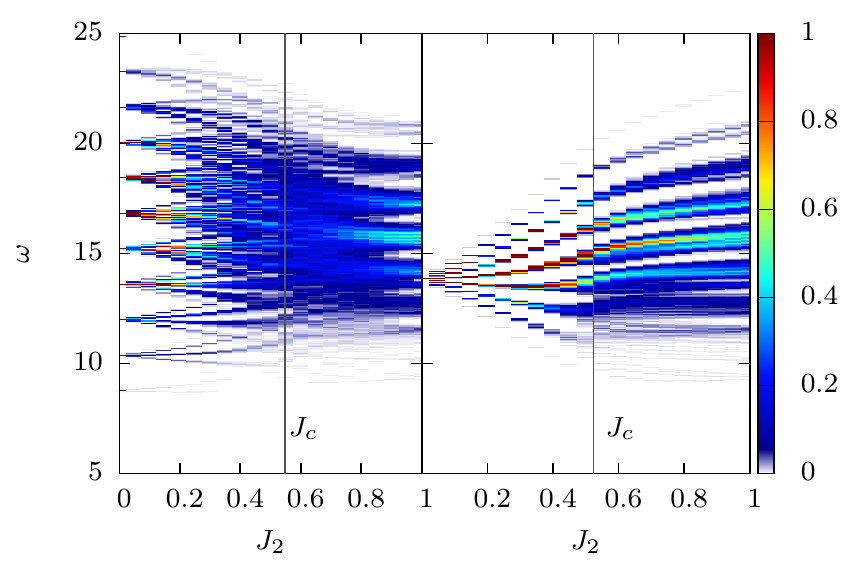} \caption{\co Development of the density
of states in the participation spectrum of $\rho_B$ accross the transition on the plaquettized
(left) and dimerized (right) square lattice.  Here, $L=20$. \label{fig:dos} } \end{figure}

Figure \ref{fig:dos} displays the density of states as a function of $J_2$ across the transition in
dimerized (right) and plaquettized (left) square lattices. While the two density of states naturally develop into the same
homogeneous limit of $J_2=1$, they appear to differ strongly for dimerized and plaquettized models, specially in the quantum disordered phase (which is nevertheless physically similar for both models).

This difference is readily understood by considering the limit of $J_2=0$ (see Appendix~\ref{sec:exact}) where the reduced density matrix for isolated plaquettes and dimers is quite different. Indeed, all diagonal elements of the reduced density matrix are equal to $1/2^L$ for the dimer case, whereas this large degeneracy is lifted by the existence of two different diagonal entries in the reduced density matrix at $J_2=0$ for a single plaquette. For the plaquettized model, the number $n_\text{strong}$ of strong domain walls determines the value of the diagonal reduced density matrix element of the line by
\begin{equation} \rho_{ii}^{{\rm line},\square} = \left(\frac{1}{12}\right)^{n_\text{strong}}
\left(\frac{5}{12}\right)^{\frac{L}{2}-n_\text{strong}}.  \end{equation}
and therefore labels the different packets of states at $J_2=0$. This  degeneracy is lifted at $J_2 > 0$ and
different packets of states tend apart from their initial pseudo-energy $\xi_i$. Due to the high
complexity of the spectrum in the plaquettized model, these packets are mixed in energy at the
critical point but become ``fat'' as they cross the critical point. This phenomenon, which will be discussed in detail in Sec. ~\ref{sec:fullspectrum}, is even more visible in the participation spectrum of the dimerized model. There, there is no distinction between strong and weak domain walls, and the well-separated bands evolve smoothly with $J_2$ across the phase transition. 

\begin{figure}[b] \includegraphics{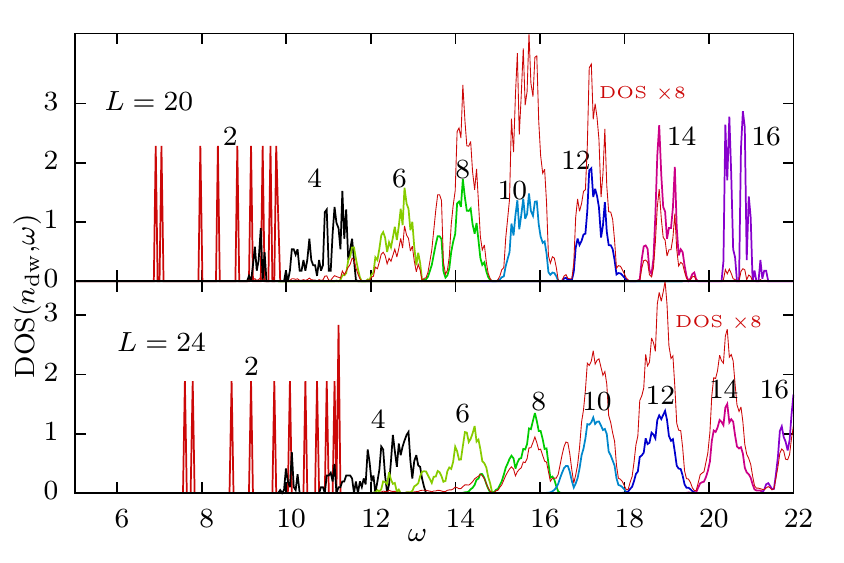} \caption{\co Density of states for different
classes of basis states, distinguished by the number of domain walls $n_{\rm dw}$ in the Heisenberg limit
$J_2=1$.  Basis states with different numbers of domain walls are clearly separated in pseudo-energy $\omega$, while the bands come closer with system size and have a small overlap. Note that the density of states is normalized for each
state band. The fine line displays the overall density of states scaled by a factor for visibility.
\label{fig:dosdw} } \end{figure}

\begin{figure} 
    \includegraphics{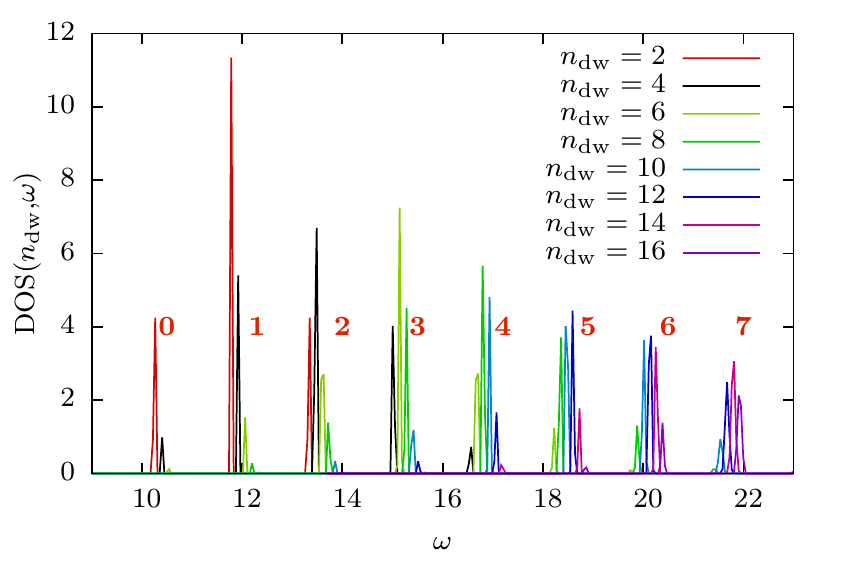} 
    \caption{\co Domain wall resolved density of states for the plaquettized model in the gapped
        phase ($J_2=0.1$ and $L=24$). The number of domain walls is not a good
identifier of the state packets in this case. In the gapped phase, even in the thermodynamic limit
there are pronounced gaps between the bands of states which in this case are characterized by the
number of strong domain walls $n_\text{strong}$ (indicated by bold face (red) numbers).  
\label{fig:dos01} } 
\end{figure}

On the other side of the transition, the density of states in the homogeneous limit $J_2=J_1$ shows bands of high density
in pseudo-energy, separated by local minima. Inspection of the corresponding states confirms our intuition by revealing that the bands can be characterized by the number of domain walls $n_{\rm dw}$ in the basis states.
Fig.~\ref{fig:dosdw} illustrates this by displaying the density of states for a fixed number of
domain walls $n_\text{dw}$ \begin{equation} \text{DOS}(n_\text{dw},\omega) =
\frac{1}{N_{n_{\text{dw}}}} \sum_{i \text{, \#dw=}{n_\text{dw}}} \delta(\xi_i - \omega),
\end{equation} where the sum runs only over states with $n_\text{dw}$ domain walls.  We compare the
domain wall resolved density of states for two lattice sizes $L=20$ and $L=24$ in Fig.~\ref{fig:dosdw} and it is apparent that the bands seen in Fig.~\ref{fig:dos} in the ordered phase
correspond to states with a fixed number of domain walls. 
This is in contrast with the gapped phase of the plaquettized model, where the number of strong domain walls is the dominant characteristic for the bands (see Fig. \ref{fig:dos01}).
With growing system size, the number of
possible bands grows linearly, as the maximal number of domain walls grows linearly in $L$. Also, on the ordered side of the transition
the width of the bands grows linearly in $L$ (as will be shown in Sec.~\ref{sec:detecting}) and the bands come closer together, eventually forming a continuum of states. 

A similar picture of bands labelled by the number of spin flips has also been proposed for the entanglement spectrum of quantum dimer models on a cylinder~\cite{stephan_renyi_2012}.

\subsection{Fully-resolved participation spectrum  resolution} \label{sec:fullspectrum}

\begin{figure}[t] \includegraphics{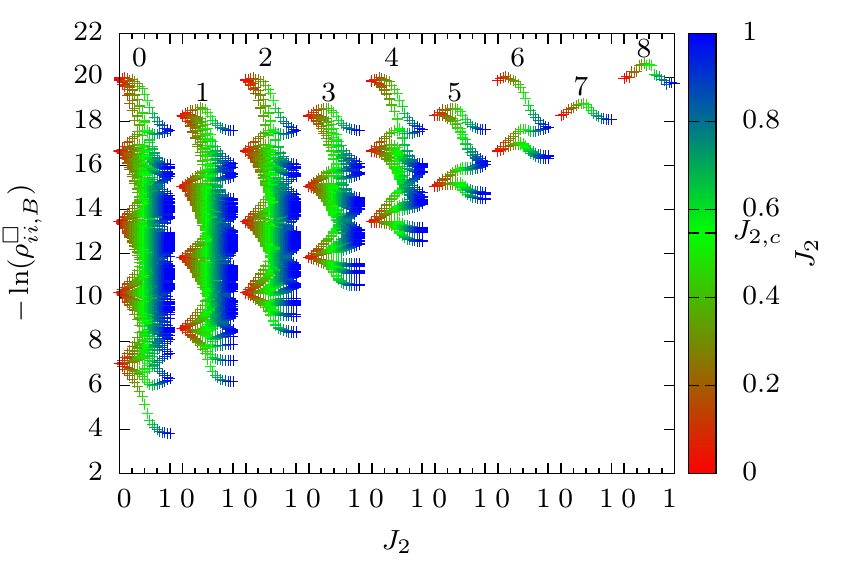} \caption{\co Development of participation
        spectrum as a function of $|S^z|$ (indicated by the numbers on top of each state column) 
        across the plaquettization transition for $L=16$.
        \label{fig:plaq_spectrum} } \end{figure}

\begin{figure} \includegraphics{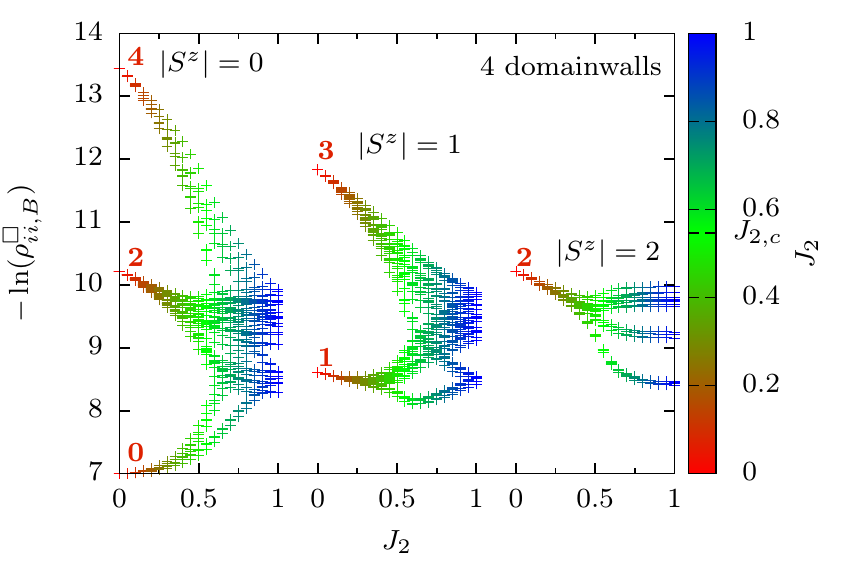} \caption{\co The band of states with 4
domain walls in the homogeneous Heisenberg limit is formed from isolated plaquette states with 0, 1,
2, 3 and 4 strong domain walls ($L=16$). Bold face (red) numbers indicate the number of strong bonds
for the state packet.  \label{fig:plaq_spectrum_dw} } \end{figure}

We now turn our attention to the fully resolved participation spectrum for the plaquettization
transition, where states are
distinguished by $(n_\text{strong},n_\text{dw},S^z)$. The use of magnetization $|S^z|$ is mainly for clarity reasons: knowing $S^z$ does not help in resolving bands which are characterized by the number of domain walls (even though some values of magnetization do not accommodate all possible number of domain walls).  The splitting of the participation spectrum in different sectors of $S^z$ is displayed in Fig.~\ref{fig:plaq_spectrum} for different values of $J_2$ for the plaquettized model, illustrating more clearly how the pseudoenergies $\xi_i$ vary from being grouped by their number of strong domain walls $n_{\rm strong}$ (at $J_2=0$) to their total number of domain walls $n_{\rm dw}$ (at $J_2=1$). A more detailed look at the development of state packets with fixed $(n_\text{strong}, n_\text{dw}, S^z)$ is provided in Fig~\ref{fig:plaq_spectrum_dw} where we concentrate on all basis states
with exactly $n_{\rm dw}=4$  domainwalls.

\subsection{Detecting the quantum phase transition in the participation spectrum}
\label{sec:detecting}

While the previous sections aimed at illustrate how participation spectra evolve though the continuous quantum phase transition, we present now two distinct quantitative features for localizing the quantum phase transition in the (fully-resolved) participation spectrum.

\subsubsection{Inflection point}
As visible in Figs.~\ref{fig:plaq_spectrum} and \ref{fig:plaq_spectrum_dw}, the behavior of the pseudo-energy $\xi_i$ of each basis state is
peculiar close to the quantum critical point, showing a maximal or minimal steepness of $\xi_i(J_2)$ and a zero transition of the second
derivative $\D^2 \xi_i/\D J_2^2$, \textit{i.e.} every $\xi_i(J_2)$ shows an inflection point at the
critical point. While the change of curvature of $\xi_i(J_2)$ close to $J_c$ is clearly visible for
most basis states in Figs. \ref{fig:plaq_spectrum} and \ref{fig:plaq_spectrum_dw}, a quantitative
demonstration that the inflection point lies right at the critical point requires high accuracy of
$\xi_i$ and a fine grid in $J_2$. We therefore performed this analysis for
$S_\infty^{\text{line}}=\xi_{\rm min}$ for best control.

\begin{figure}[t]
    \centering
    \includegraphics{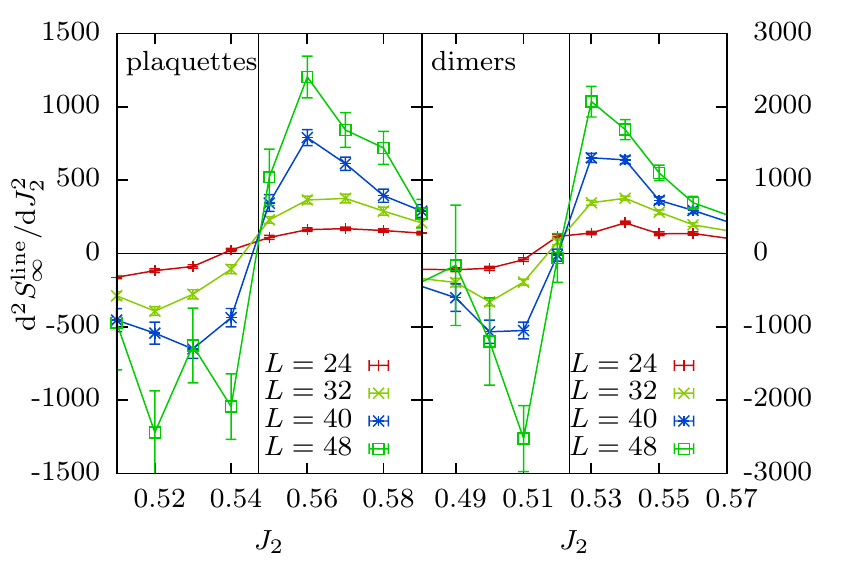}
    \caption{\co Second derivative of $S_\infty^{\text{line}}$ for the plaquettized (left) and
    dimerized (right) lattices for different lattice sizes.
The sign change shows a clear inflection point right at the quantum phase transition, which
corresponds to a pronounced minimum of the first derivative of $S_\infty^\text{line}$.
}
    \label{fig:plaq_line_sinf_diff2}
\end{figure}

We compute numerically the second derivative of $S_\infty^{\text{line}}$ with respect to $J_2$ close
to the quantum phase transition and display it in Fig. \ref{fig:plaq_line_sinf_diff2}  for both the dimerized and plaquettized models, confirming that the quantum phase transition corresponds to a zero in the second derivative. In the limit $L\rightarrow \infty$,
the first derivative of $S_\infty^\text{line}$ diverges right at the critical point and thus precisely marks the quantum phase transition.

\subsubsection{Finite-size dependence of resolved bands and domain walls confinement}
\begin{figure}[h]
    \includegraphics{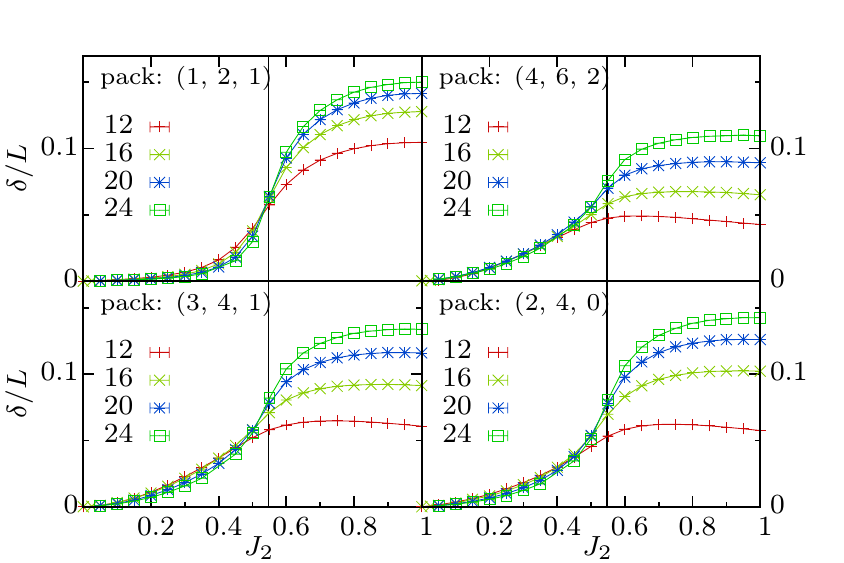}
 \includegraphics{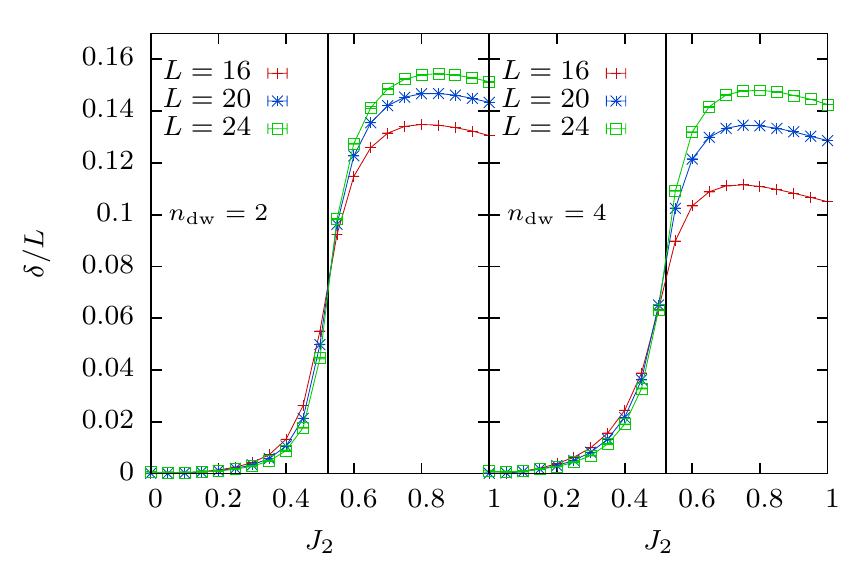}
\caption{\co 
Dependence of the width $\delta/L$ of state
packets as a function of system size $L$ for fixed $(n_\text{strong},n_\text{dw},|S^z|)$ (top panel)
and $n_\text{dw}$ (bottom panel, including states from all $S^z$ sectors) across the plaquettization (top) or dimerization (bottom) transition. Lower numbers of domain walls $n_\text{dw}$ correspond to low pseudo-energy part of the participation spectrum. The crossing of $\delta/L$ for different sizes close to the
critical point indicated by a vertical line is most prominent for the subsystem basis state packets
with low pseudo-energy (\textit{e.g.} for the packet with $2$ domain walls for both panels) but clearly exists with
considerable larger size effects (such as drift crossing) in higher parts of the spectrum.  \label{fig:allpacks}}
\end{figure}

\begin{figure}
    \includegraphics{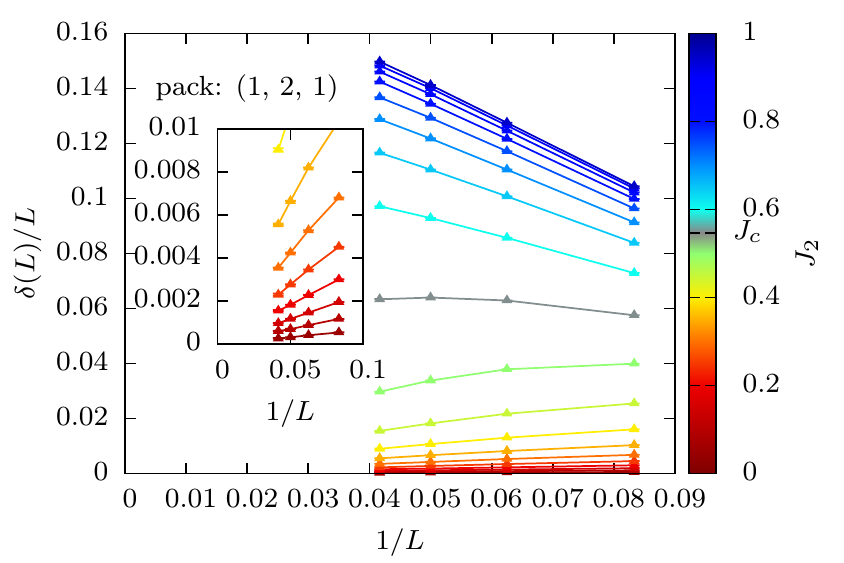}
    \caption{\co
    Width $\delta/L$ of state packet (1, 2, 1) as a function of inverse system size $1/L$ 
        for different values of $J_2$ in the case of the plaquettization transition. In the limit of
        large system sizes, $\delta/L$ vanishes in the quantum disordered phase, while it goes to a
        constant in the ordered phase. Inset: Zoom for small values of $J_2$ in the gapped phase.
        The linear behavior in $1/L$ suggests that $\delta/L$ vanishes like $1/L$.
        \label{fig:deltaL}}
\end{figure}

Another way to quantitatively detect the quantum phase transition is obtained following the previous observation that bands of identical number of domain walls in the participation spectra appear to become ``fatter" as the system crosses the quantum critical point.

We investigate the size of state packets labeled by $P=(n_\text{strong}, n_\text{dw},|S^z|)$ in the
plaquettized case by looking at the difference in pseudo-energy $\delta=\xi_i^{\rm
max}(P)-\xi_i^{\rm min}(P)$ between the basis states with the largest and lowest pseudo-energy
within the packet $P$. Results for the normalized packet width $\delta / L$ for different $P$ (see
top panel of Fig.~\ref{fig:allpacks})  display a clear crossing point at the quantum critical point, separating
two different regimes. In the quantum disordered phase, the normalized packet width $\delta / L$
tends to vanish presumably as $1/L$ in the thermodynamic limit, as revealed by the finite-size
scaling analysis in the inset of  
Fig.~\ref{fig:deltaL}. On the other hand, the packet width $\delta$ grows as $L$ (with $1/L$
correction) in the ordered phase as also seen in Fig.~\ref{fig:deltaL}.
  The same behavior is revealed when integrating over all $S^z$ sectors as shown for the case of the
  dimerized model in the bottom panel of Fig.~\ref{fig:allpacks}, where state packets are defined by
  a fixed number of domain walls $n_\text{dw}$.

\begin{figure}[h]
    \centering
    \includegraphics{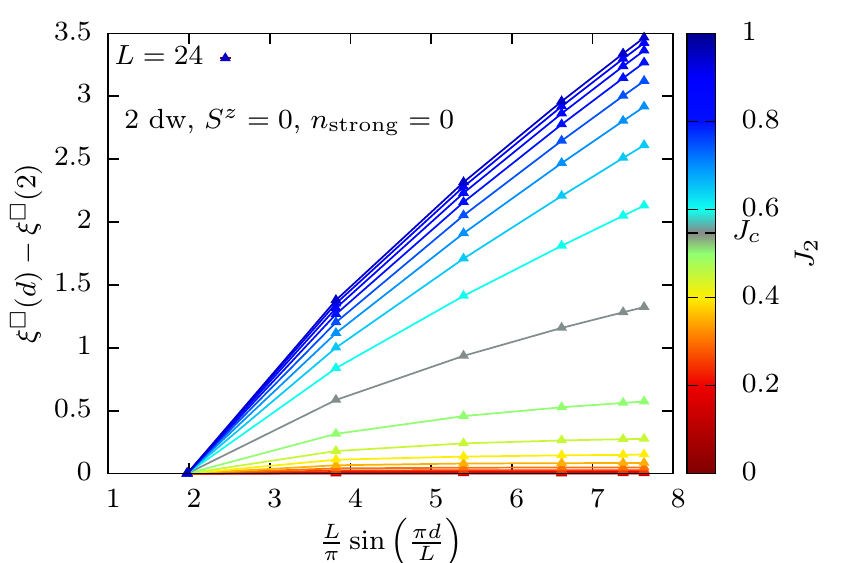}
    \includegraphics{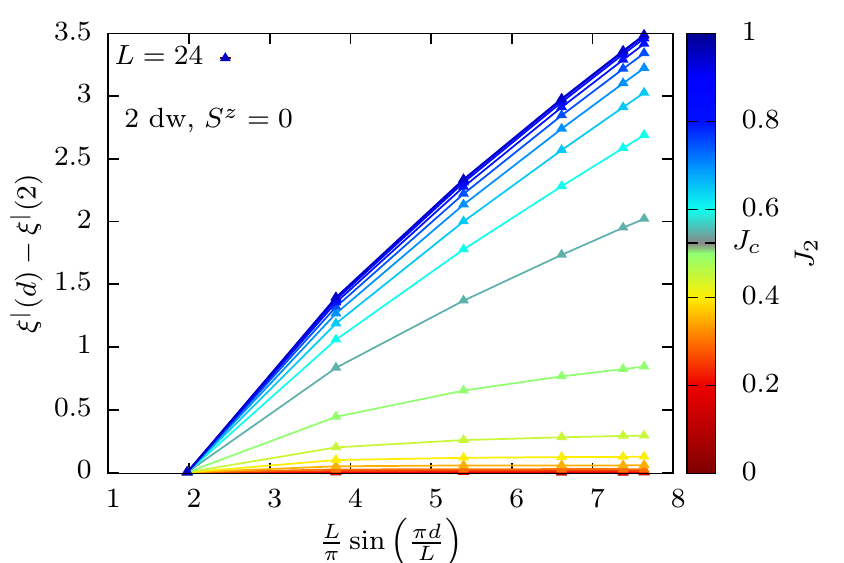}
    \caption{
        \co
        Pseudo-energies of the line subsystem as a function of domain wall distance $d$ (expressed in
        terms of the chord distance on the ring) for all states with $2$ domain walls in the $S^z=0$
        sector for the plaquettized (top) and dimerized (bottom) model, for different values of
        $J_2$ (equidistant in steps of $0.05$).  In order to compare the curves, we subtracted the pseudo-energy of the state with the minimal distance $d=2$ between domain walls (which is always the lowest).
}
    \label{fig:dwpotential}
\end{figure}

This behavior can be understood following the sketch presented in Fig.~\ref{fig:dwspict}, where one sees that the dynamics of two domain walls is constrained by the relative sizes of two N\'eel patterns, $\rm N_A$ and $\rm N_B$. Since the staggered magnetization of a single basis state depends on the size difference $|\ell_{\rm A}-\ell_{\rm B}|$ between N\'eel domains of different type, one can predict a qualitative difference for the packet width $\delta$ between ordered and disordered ground states.
 In the disordered phase $J_2<J_c$ where antiferromagnetic correlations are short-ranged, we naturally expect that the pseudo-energies will not be affected (remaining essentially constant) when the separation between two domain walls becomes larger than the finite correlation length. On the other hand,  for long-range order a confinement mechanism between two domain walls will be necessary to maintain a finite staggered magnetization. There we expect the pseudo-energy spectrum to be controlled by an attractive long-range interaction between domain walls.  
 This picture holds best for a small number of domain walls in the high probability (low pseudo-energy) part of the
participation spectrum.

We illustrate this interpretation by considering the pseudo-energy dependence on the distance between domain walls. Fig.~\ref{fig:dwpotential} displays this pseudo-energy difference for the diluted case of $n_{\rm dw}=2$ domain-walls with $S^z=0$ (in which case the distance has to be even), as a function of the chord distance between domain walls for different $J_2$ for both dimerized and plaquettized models. In the disordered phase, the domain walls appear rapidly deconfined with a finite small pseudo-energy difference between states with
different domain wall distances and hence a small packet width $\delta$. In contrast, large domain wall distances are penalized in the magnetically ordered phase by a high pseudo-energy cost, which appears to grow approximatively linearly with distance (for a large enough distance).

For states with more than two domain walls, the situation becomes somewhat more complicated. The crossing in $\delta/L$  
still exists but it acquires a drift with system size (see Fig.~\ref{fig:allpacks}). A closer inspection of the involved states in
the corresponding packet and their pseudo-energies suggests that multi-domain wall attraction terms play a
role in addition to the long-range $2$ domain-walls attraction. This is probably the source of the a drift of the crossing as
these multi-domain wall terms become eventually less important when the average distance between domain walls becomes large
$L\gg n_\text{dw}$ for the high end (low probability) of a packet with a fixed number of domain walls.

\section{Conclusion}
\label{sec:conclusion}

We have analyzed the N\'eel antiferromagnet-paramagnet quantum phase transition in two-dimensional quantum spin systems using the Shannon-R\'enyi entropies for the full system and for subsystems,
together with the associated participation spectra, using extensively QMC methods presented in
Ref.~\onlinecite{luitz_universal_2014}. Our study shows that a line shaped one-dimensional subsystem
is actually sufficient to capture the quantum phase transition. We confirm that the
subleading scaling behavior of Shannon-R\'enyi entropies changes radically at the quantum phase transition, giving rise to a logarithmic scaling term in the N\'eel phase, independent on microscopic details (such as the choice of dimerization or plaquettization of the lattice in our study). 

Similar logarithmic corrections to an {\it area law} have been numerically observed\cite{kallin_anomalies_2011,humeniuk_quantum_2012,ju_entanglement_2012} in studies of the R\'enyi entanglement entropy of the ground-state of the 2d Heisenberg model for
{\it two-dimensional} subsystems (such as half-torus), in agreement with theoretical predictions~\cite{metlitski_entanglement_2011}.
For the case of a line-shaped subsystem, we have checked that the R\'enyi entanglement entropy also exhibits a
logarithmic correction to the area law (indistinguishable from a volume law in this particular case).

On the disordered side of the phase diagram, only a subleading constant term can be present, which is actually $0$ in this phase with no broken symmetry.

At the quantum critical point, we find a universal subleading constant (for the line SR entropy $S_\infty^{\rm line}$) with an estimated value of $b_\infty^{*,\text{line}}=0.41(1)$ being identical for dimerized and plaquettized models. We have confirmed that a similar constant can be found at the N\'eel-paramagnetic finite-temperature phase transition of the 3d Heisenberg $S=1/2$ antiferromagnet at $T_c/J=0.94408(2)$, strongly suggesting that this is a characteristic of the 3d $O(3)$ universality class to which all mentioned transitions belong. We suspect that such a universal subleading constant also exists for the SR entropy of the full system, but the precision of our numerical computations do not allow to prove this. Universality could be further checked by considering $O(3)$ critical points in other models, either with two-dimensional quantum models such as $S=1/2$ bilayers~\cite{Sandvik94}, coupled Haldane chains~\cite{Sakai89,matsumoto_ground-state_2001}, staggered-dimer models where anomalously large corrections to scaling are known~\cite{Fritz11}, or in 3d classical systems such as the classical Heisenberg model on the cubic lattice. It would also be very interesting to extend this study to other universality classes, either of conventional or unconventional type~\cite{Senthil04}.

Motivated by the finding that the one-dimensional subsystem captures the physics of the phase
transition, we also performed a phenomenological study of the information contained in the participation
spectra of the subsystem. In the participation spectra, states group together in packets of pseudo-energy which can be classified by the number of ferromagnetic domain walls (number of strong domain walls for the plaquettized lattice) in the basis states. The development of these packets across the transition is peculiar
as packets become ``fat'' in the ordered phase, meaning that their width $\delta$ (in pseudo-energy) grows linearly with subsystem size $L$. In the disordered phase, however, $\delta/L$ tends to zero. This can be phenomenologically explained by an attractive potential between domain walls in the ordered phase, while in the disordered phase, domain walls remain deconfined. In addition, we observe at the critical point an interesting behavior for the pseudo-energies in the participation spectrum: their slope as a function of the control parameter $J_2$ has a pronounced extremum and we show that $S_\infty^\text{line}$ has an inflection point at $J_c$. 

The analysis presented in Sec.~\ref{sec:PS} is fairly simple, but contains the basic ingredients to build a wave-function that describes the studied (2+1)-dimensional quantum phase transition. Let us indeed put in perspective our work with the widely-used variational approach. In the context
of two-dimensional Heisenberg models, Huse and Elser \cite{huse_simple_1988} (see also Refs.~\onlinecite{manousakis_paired_1989,liu_variational_1989}) formulated a variational ansatz for the ground-state wave function:
\begin{equation}
    \ket{\psi}({\bf \alpha}) =  \sum_i \E^{-\frac{\xi_i({\bf \alpha})}{2}} \ket{i},
    \label{eq:var}
\end{equation}
where a classical pseudo-energy $\xi_i$ is associated to the basis state $\ket{i}$ (also taken as a
$\{S^z\}$ basis state in Ref.~\onlinecite{huse_simple_1988}). Here ${\bf \alpha}$ is (a set of)
variational parameter(s) used to minimize the total energy of the Heisenberg Hamiltonian. This
ansatz looks similar to our definition of the participation spectrum. The crucial difference is of
course that the variational approach {\it assumes} a form for the pseudo-energy (for instance a power-law Ising interaction in Ref.~\onlinecite{huse_simple_1988}), while our QMC
methods can calculate the {\it exact} (within statistical accuracy) value of each $\xi_i$. Note as
well that we did not compute, for essentially practical reasons, the participation
spectrum for the full system (as in the variational ansatz) but rather on a subsystem. Nevertheless,
our results give the correct qualitative ingredients to construct variational wave-functions of the form of
Eq.~\eqref{eq:var} to describe the N\'eel and paramagnetic phases, as well as the transition in between. This could be useful in particular for frustrated spin systems, where QMC is not available. 

\begin{acknowledgments}
    We would like to thank F. Assaad, F. B\`egue, G. Misguich, G. Roux, S. Pujari and J.-M.
    St\'ephan for many useful discussions, as well as G. Misguich, M. Oshikawa and X. Plat for collaboration on
    related topics.
    Our QMC codes are partly based on the ALPS
    libraries~\cite{bauer_alps_2011}. This work was performed using numerical resources from GENCI (grants 2013-x2012050225 and 2014-x2013050225) and CALMIP and is supported by the French ANR program ANR-11-IS04-005-01.
\end{acknowledgments}

\appendix 

\section{Details on Monte Carlo procedure} \label{sec:MC}

Here we provide further details (in supplement to Ref. \onlinecite{luitz_universal_2014}) on the QMC procedure
to access subsystem entropies and participation probabilities, as well as on dealing with
statistical uncertainty.

\subsection{Calculation of diagonal elements of the reduced density matrix}

  As already presented in Ref. \onlinecite{luitz_universal_2014}, the diagonal elements $\rho_{ii}$
  of the full density matrix are readily available in standard QMC techniques such as the standard stochastic series expansion
  method (cf. \textit{e.g.} Ref. \onlinecite{sandvik_computational_2010}). In the same vein, the
  diagonal elements of the reduced density matrix $\rho_{B,ii}$ (see Eq.~\eqref{eq:rhob}) are obtained
  in the following way. It is clear that the probability of observing the basis state $\ket{j(i_B)}$
  in the SSE operator string is directly given by \begin{equation} p(\ket{j(i_B)}) = \rho_{j(i_B)
  j(i_B)}.  \end{equation} Hence, the probability of finding the subsystem basis state $\ket{i_B}_B$
  in subsystem $B$ in the stochastic series expansion operator string is \begin{equation} p(\ket{i_B}_B) = \sum_{j(i_B)}
  \rho_{j(i_B) j(i_B)} = \rho_{B,ii}.  \end{equation}

In practice, this means that one has to inspect only the $B$-part of each basis state in the simulation,
and record a histogram of its frequency.

\subsection{Computationally accessible entropies}

The numerical calculation of large SR entropies corresponds to the observation of rare events with low probabilities.
It is easy to obtain an approximation of a reasonable upper limit for the calculation of the entropy
$S_\infty=-\ln p_\text{max}$ given by the observation of the most probable state appearing in the SSE
Markov chain with a probability $p_\text{max}$. For the estimation of the maximally accessible
entropy, we make the assumption of statistical independence, as statistical correlations will only
reduce the maximally accessible entropy. 

In the Markov chain, we measure the observable $\delta_{i,i_\text{max}}$, yielding $1$ of state $\ket{i}$
corresponds to the state $\ket{i}_\text{max}$ with maximal probability and $0$ otherwise. The Monte Carlo average of this observable
is given by

\begin{equation} 
    \langle \delta_{i,i_\text{max}} \rangle = \sum_{s \in \text{MC}}
    \delta_{s,i_\text{max}} = \frac{n(i_\text{max})}{N_{\text{MC}}}, 
\end{equation}

where $n(i_\text{max})$ is the number of occurrences of the most probable state in the Markov chain
$\text{MC}$ of
length \footnote{Note that in the case of SSE, the effective length of the Markov chain is rather the
number of operator strings $N_s$ times the expansion order $n$.} $N_\text{MC}$.

The standard error of $\langle \delta_{i,i_\text{max}} \rangle$ is given by 
\begin{equation} 
    \sigma_p =
    \frac{\sqrt{p_\text{max} (1-p_\text{max})}} {\sqrt{N_\text{MC}}} \approx \sqrt {
        \frac{p_\text{max}}{N_\text{MC}} } 
\end{equation}
for sufficiently small $p_\text{max}$ and $N_\text{MC}$ inspected states. Now we can estimate the standard error
of $S_\infty$ by the linear approximation $\sigma_S = \sigma_p/p$. To obtain a given relative error
$\sigma_S/S_\infty$ for a given number $N_\text{MC}$ of inspected states, the maximal entropy is then governed
by the equation:

\begin{equation} \frac{\sigma_S}{S_\infty} S_\infty = \frac{1}{\sqrt{N_\text{MC}}}
\frac{1}{\mathrm{e}^{-S_\infty/2} }.  \end{equation}

In a realistic calculation, we inspect typically $N_\text{MC}=10^{12}$ states. Therefore, for a requirement of
a relative error $\sigma_S/S_\infty \leq 0.001$, we obtain a maximally accessible entropy
$S_\infty \leq 19.78 \lesssim 20$.

\subsection{Error bars for the participation spectrum}

  Measuring accuracy on the participation spectrum requires special care. It is indeed impractical to construct an observable for each subsystem
  basis state $\ket{i_B}$ and measure $\delta_{i_B,j_B}$ as proposed in Ref.
  \onlinecite{luitz_universal_2014} for each observed state $\ket{j_B}$ in the SSE operator string. For the
  calculation of the participation spectrum, defined as $\xi_{i_B} = -\ln \rho_{B,i_B i_B}$ and inducing a nonlinear transformation of the Monte Carlo data,
we perform several independent Monte Carlo simulations in parallel, each creating a simple
histogram $h(\ket{i_B})$ containing the count of all observed subsystem basis states. After applying
all model symmetries to the histogram to improve the statistics (see supplementary material of Ref.
\onlinecite{luitz_universal_2014}), we perform a bootstrap analysis for each $\xi_{i_B}$, creating
bootstrap samples from the histogram counts of the different independent simulations. This provides
an unbiased estimation of the standard errors of the mean of all $\xi_{i_B}$.

We did not display error bars in Figures \ref{fig:plaq_spectrum} and \ref{fig:plaq_spectrum_dw} as
they are smaller than the line width but they where used for the estimation of the uncertainty in
$\delta/L$ shown \eg in Fig. \ref{fig:deltaL}.

\section{Exact calculations for $J_2=0$}
\label{sec:exact}

We provide here simple exact results for SR entropies and reduced density matrices when $J_2=0$. These  results are useful
in two ways: they provide direct insight for the subleading terms of the
SR entropies (which vanish altogether when $J_2=0$) as well as on the starting point of the participation spectrum. 

\subsection{Shannon R\'enyi entropy of the full system}

In the limit $J_2=0$ of isolated plaquettes or dimers, the groundstate
$\ket{\varphi}$ is a singlet state,
given by \begin{equation} \begin{split} \ket{\varphi}_\square &= \frac{1}{\sqrt{12}} \left( -2
\ket{\!\begin{array}{c}\dw \up \\ \up \dw\end{array}\!} -2 \ket{\!\begin{array}{c}\up \dw \\ \dw
\up\end{array}\!}\right. \\ &\left. +\ket{\!\begin{array}{c}\dw \dw \\ \up \up\end{array}\!} +
\ket{\!\begin{array}{c}\up \dw \\ \up \dw\end{array}\!} + \ket{\!\begin{array}{c}\dw \up \\ \dw
\up\end{array}\!} + \ket{\!\begin{array}{c}\up \up \\ \dw \dw\end{array}\!} \right) \end{split}
\end{equation} for one plaquette (see {\it e.g.} Ref.~\onlinecite{ueda_ground-state_2007})) and \begin{equation} \ket{\varphi}_| = \frac{1}{\sqrt{2}} \left(
\ket{\up\dw} - \ket{\dw\up} \right) \end{equation} for a single dimer.

Thus, we get \begin{equation} S_\infty^\square = - \ln \left( \frac{4}{12} \right)^{N/4} = \frac{\ln
3}{4} N.  \end{equation}

and \begin{equation} S_q^\square = \frac{1}{1-q} \frac{N}{4} \ln \left[ 4 \left( \frac 1 {12}
\right)^q + 2 \left( \frac 4 {12} \right)^q \right] \end{equation}
for the limit $J_2=0$ of the plaquettized model. 

The solution is much simpler for the dimerized model, as any basis state with $S^z=0$ on
the dimer contributes with the same weight $p_i=2^{-N/2}$ and therefore all SR entropies are
identical: \begin{equation} S_q^| =  \frac{\ln 2}{2} N.  \end{equation}

\subsection{Shannon R\'enyi entropy of the line sub-system} 

For a dimer shaped subsystem $B$ of one
plaquette, the reduced density matrix is given 
(in the $\{ \fullket{\up \up}, \fullket{\up \dw}, \fullket{\dw \up}, \fullket{\dw \dw} \}$ basis) by \begin{equation} \rho_B^{\square} =
\mathrm{Tr}_{\text{A}} \ket{\varphi}\bra{\varphi}= \frac{1}{12} \begin{pmatrix} 1 & 0 & 0 & 0 \\ 0 & 5
& -4& 0 \\ 0 & -4& 5 & 0 \\ 0 & 0 & 0 & 1 \end{pmatrix}.  \end{equation} In the single dimer case,
subsystem $B$ is just one site and the reduced density matrix (in the $\{ \fullket{\up}, \fullket{\dw}\}$ basis) is
\begin{equation} \rho_B^{|} = \frac{1}{2} \begin{pmatrix} 1 & 0 \\ 0 & 1 \end{pmatrix}.
\end{equation}

The total reduced density matrix of the line shaped subsystem composed of $L/2$ dimers $B$ of single
plaquettes (or $L$ sites $B$ of single dimers in the dimerized case) is then obtained from the
Kronecker product of the reduced density matrices:

\begin{equation} \rho^{\text{line},\square} = \bigotimes \limits_{i=1}^{L/2}
\rho_{i,\text{B}}^\square \quad \text{or} \quad \rho^{\text{line},|} = \bigotimes \limits_{i=1}^{L}
\rho_{i,\text{B}}^|.  \end{equation}

Consequently, for the line shaped subsystem, $S_\infty^{\rm line}$ reduces to \begin{equation}
S_\infty^{\text{line},\square} = L \frac{ \ln \frac{12}{5} }{2} \quad \text{or} \quad
S_\infty^{\text{line}, |} = L \ln 2.  \end{equation}

While for arbitrary values of $q$, the Shannon-R\'enyi entropies are \begin{equation}
S_q^{\text{line},\square} = \frac{L}{2} \frac{ \ln\left[ 2 \left( \frac{1}{12} \right)^q + 2 \left(
\frac{5}{12} \right)^q \right] } {1-q} \quad \text{or} \quad S_q^{\text{line},|} = L \ln 2.
\end{equation}

\end{document}